\documentclass[%
 twocolumn,
 amsmath,amssymb,
 aps,
 prc,
 superscriptaddress,
 showkeys,
 showpacs,
 preprintnumbers,
 floatfix
]{revtex4}

\usepackage{comment}

\usepackage{cancel}
\usepackage[colorlinks]{hyperref}
\hypersetup{%
    ,urlcolor=blue
    ,citecolor=blue
    ,linkcolor=blue
    }

\usepackage{xcolor}
\definecolor{magenta}{rgb}{0.8, 0.0, 0.8}

\usepackage{graphicx}% Include figure files
\usepackage{dcolumn}% Align table columns on decimal point
\usepackage{bm}% bold math
\usepackage[mathlines]{lineno}% Enable numbering of text and display math
\usepackage[mathlines]{lineno}% Enable numbering of text and display math
\usepackage{placeins}
\usepackage{multirow}
\usepackage[normalem]{ulem}
\usepackage{lineno}
\newcommand{\ppt}{p_{\rm T}}
\newcommand{\sqrtsNN}{\mbox{$\sqrt{s_{\mathrm{NN}}}$}}

\def \auau {\mbox{Au+Au}}
\newcommand{\mean}[1]{\left\langle #1 \right\rangle}

\normalsize % necessary for booktabs package
\usepackage{booktabs}
\usepackage{appendix}
\usepackage{listings}
\UseRawInputEncoding

\begin{document}

\title{Higher-Order Cumulants and Correlation Functions of Proton Multiplicity Distributions in $\sqrtsNN$ = 3 GeV \auau{} Collisions at the RHIC STAR Experiment}

\affiliation{American University of Cairo, New Cairo 11835, New Cairo, Egypt}
\affiliation{Texas A\&M University, College Station, Texas 77843}
\affiliation{Czech Technical University in Prague, FNSPE, Prague 115 19, Czech Republic}
\affiliation{AGH University of Science and Technology, FPACS, Cracow 30-059, Poland}
\affiliation{Ohio State University, Columbus, Ohio 43210}
\affiliation{University of Kentucky, Lexington, Kentucky 40506-0055}
\affiliation{Panjab University, Chandigarh 160014, India}
\affiliation{Variable Energy Cyclotron Centre, Kolkata 700064, India}
\affiliation{Brookhaven National Laboratory, Upton, New York 11973}
\affiliation{Abilene Christian University, Abilene, Texas   79699}
\affiliation{Instituto de Alta Investigaci\'on, Universidad de Tarapac\'a, Arica 1000000, Chile}
\affiliation{University of California, Riverside, California 92521}
\affiliation{University of Houston, Houston, Texas 77204}
\affiliation{University of Jammu, Jammu 180001, India}
\affiliation{State University of New York, Stony Brook, New York 11794}
\affiliation{Nuclear Physics Institute of the CAS, Rez 250 68, Czech Republic}
\affiliation{Shanghai Institute of Applied Physics, Chinese Academy of Sciences, Shanghai 201800}
\affiliation{Yale University, New Haven, Connecticut 06520}
\affiliation{University of California, Davis, California 95616}
\affiliation{Lawrence Berkeley National Laboratory, Berkeley, California 94720}
\affiliation{University of California, Los Angeles, California 90095}
\affiliation{Indiana University, Bloomington, Indiana 47408}
\affiliation{Warsaw University of Technology, Warsaw 00-661, Poland}
\affiliation{Shandong University, Qingdao, Shandong 266237}
\affiliation{Fudan University, Shanghai, 200433 }
\affiliation{University of Science and Technology of China, Hefei, Anhui 230026}
\affiliation{Tsinghua University, Beijing 100084}
\affiliation{University of California, Berkeley, California 94720}
\affiliation{ELTE E\"otv\"os Lor\'and University, Budapest, Hungary H-1117}
\affiliation{University of Heidelberg, Heidelberg 69120, Germany }
\affiliation{Wayne State University, Detroit, Michigan 48201}
\affiliation{Indian Institute of Science Education and Research (IISER), Berhampur 760010 , India}
\affiliation{Kent State University, Kent, Ohio 44242}
\affiliation{Rice University, Houston, Texas 77251}
\affiliation{University of Tsukuba, Tsukuba, Ibaraki 305-8571, Japan}
\affiliation{University of Illinois at Chicago, Chicago, Illinois 60607}
\affiliation{Lehigh University, Bethlehem, Pennsylvania 18015}
\affiliation{University of Calabria \& INFN-Cosenza, Italy}
\affiliation{National Cheng Kung University, Tainan 70101 }
\affiliation{Purdue University, West Lafayette, Indiana 47907}
\affiliation{Southern Connecticut State University, New Haven, Connecticut 06515}
\affiliation{Central China Normal University, Wuhan, Hubei 430079 }
\affiliation{Technische Universit\"at Darmstadt, Darmstadt 64289, Germany}
\affiliation{Temple University, Philadelphia, Pennsylvania 19122}
\affiliation{Valparaiso University, Valparaiso, Indiana 46383}
\affiliation{Indian Institute of Science Education and Research (IISER) Tirupati, Tirupati 517507, India}
\affiliation{Institute of Modern Physics, Chinese Academy of Sciences, Lanzhou, Gansu 730000 }
\affiliation{Frankfurt Institute for Advanced Studies FIAS, Frankfurt 60438, Germany}
\affiliation{National Institute of Science Education and Research, HBNI, Jatni 752050, India}
\affiliation{University of Texas, Austin, Texas 78712}
\affiliation{Rutgers University, Piscataway, New Jersey 08854}
\affiliation{Institute of Nuclear Physics PAN, Cracow 31-342, Poland}
\affiliation{Max-Planck-Institut f\"ur Physik, Munich 80805, Germany}
\affiliation{Creighton University, Omaha, Nebraska 68178}
\affiliation{Indian Institute Technology, Patna, Bihar 801106, India}
\affiliation{Ball State University, Muncie, Indiana, 47306}
\affiliation{Universidade de S\~ao Paulo, S\~ao Paulo, Brazil 05314-970}
\affiliation{Huzhou University, Huzhou, Zhejiang  313000}
\affiliation{Michigan State University, East Lansing, Michigan 48824}
\affiliation{Argonne National Laboratory, Argonne, Illinois 60439}
\affiliation{United States Naval Academy, Annapolis, Maryland 21402}
\affiliation{South China Normal University, Guangzhou, Guangdong 510631}

\author{M.~S.~Abdallah}\affiliation{American University of Cairo, New Cairo 11835, New Cairo, Egypt}
\author{B.~E.~Aboona}\affiliation{Texas A\&M University, College Station, Texas 77843}
\author{J.~Adam}\affiliation{Czech Technical University in Prague, FNSPE, Prague 115 19, Czech Republic}
\author{L.~Adamczyk}\affiliation{AGH University of Science and Technology, FPACS, Cracow 30-059, Poland}
\author{J.~R.~Adams}\affiliation{Ohio State University, Columbus, Ohio 43210}
\author{J.~K.~Adkins}\affiliation{University of Kentucky, Lexington, Kentucky 40506-0055}
\author{I.~Aggarwal}\affiliation{Panjab University, Chandigarh 160014, India}
\author{M.~M.~Aggarwal}\affiliation{Panjab University, Chandigarh 160014, India}
\author{Z.~Ahammed}\affiliation{Variable Energy Cyclotron Centre, Kolkata 700064, India}
\author{D.~M.~Anderson}\affiliation{Texas A\&M University, College Station, Texas 77843}
% ...

\collaboration{STAR Collaboration}\noaffiliation

\date{\today}

\begin{abstract}
We report a measurement of cumulants and correlation functions of event-by-event proton multiplicity distributions from fixed-target \auau{} collisions at $\sqrtsNN$ = 3 GeV measured by the STAR experiment. Protons are identified within the rapidity ($y$) and transverse momentum ($p_{\rm T}$) region $-0.9 < y<0$ and $0.4 < \ppt <2.0 $ GeV/$c$ in the center-of-mass frame.
A systematic analysis of the proton cumulants and correlation functions up to sixth-order as well as the corresponding ratios as a function of the collision centrality, $p_{\rm T}$, and $y$ are presented. The effect of pileup and initial volume fluctuations on these observables and the respective corrections are discussed in detail. The results are compared to calculations from the hadronic transport UrQMD model as well as a hydrodynamic model. In the most central 5\% collisions, the value of proton cumulant ratio $C_4/C_2$ is negative, drastically different from the values observed in \auau{} collisions at higher energies. Compared to model calculations including Lattice QCD, a hadronic transport model, and a hydrodynamic model, the strong suppression in the ratio of $C_4/C_2$ at 3 GeV Au+Au collisions indicates an energy regime dominated by hadronic interactions. 
\end{abstract}

\maketitle

\section{Introduction}

One of the main goals of the Beam Energy Scan (BES) program at the BNL Relativistic Heavy Ion Collider (RHIC) is to study the nature of the QCD phase diagram in a two-dimensional phase space spanned by temperature and baryonic chemical potential ($\mu_B$).
Experimental data from RHIC and the Large Hadron Collider (LHC) in collision energies where $\mu_B$ approaches zero, have provided evidence of a quark-gluon plasma (QGP)~\cite{BRAHMS:2004adc,PHOBOS:2004zne,PHENIX:2004vcz,STAR:2005gfr}. In this region where \mbox{$\mu_B\sim 0$ MeV}, lattice QCD (LQCD) predicts a smooth crossover from a QGP phase to a hadronic state~\cite{Borsanyi:2013bia,Gupta:2011wh}. The QGP matter has been found to hadronize at temperatures close to the lattice QCD estimated transition temperature at \mbox{$\mu_{\rm B}=0$ MeV}~\cite{Borsanyi:2010bp,HotQCD:2018pds}.

In the region where $\mu_B$ is finite, the nature of the transition to QGP matter is less understood. Various models favor a first-order phase transition~\cite{Halasz:1998qr}, which requires the existence of a critical endpoint.
Ideally, near the critical point, the correlation length could grow. Provided that the signal of the critical point develops as fast as the system expands, the critical point could be experimentally measured.
The higher-order event-by-event fluctuations of conserved quantities such as net charge, net baryon, and net strangeness are expected to be sensitive to the correlation length $\xi$, and thus may serve as indicators of critical behavior~\cite{PhysRevLett.91.102003,2017overview,2017FC_Kitazawa,Asakawa:2015ybt}. A general expectation of the critical-point-induced fluctuations is that the net-baryon higher-order cumulant ratios (e.g. $C_4/C_2$) oscillate with collision energy~\cite{Stephanov:2011pb,Stephanov:2008qz,Stephanov:2011zz}. In heavy-ion collisions, however, effects of finite size and limited lifetime of the hot nuclear system may put constraints on the significance of signals~\cite{FRAGA,2017ProtonFC_UrQMD, PhysRevC.98.054620}. Here, cumulants are a set of quantities which provide an alternative to moments of a probability distribution. Their definitions can be found at Sec.~\ref{cum:def}.

At small $\mu_{\rm B}$, LQCD calculations have predicted positive cumulant ratios of $C_4/C_2$ and negative ratios of $C_5/C_1$ and $C_6/C_2$ in the regime where the QGP is expected to exist. The results suggest that a critical point below \mbox{$\mu_{\rm B}<200$ MeV} is unlikely~\cite{STAR:2010mib}. The first phase of the RHIC Beam Energy Scan program (BES-I), conducted in \mbox{2010 -- 2014}, covered energies from \mbox{$\sqrtsNN=7.7$ GeV} to \mbox{$\sqrtsNN=200$ GeV} and generated several results on directed and elliptic flows which suggest a change in the equation of state of QCD matter~\cite{STAR:2013cow,STAR:2017ieb,STAR:2014clz}. Recently, a study from BES-I~\cite{Adam:2020unf,Abdallah:2021fzj} has shown a non-monotonic behavior of the cumulant ratio $C_4/C_2$ of the net-proton multiplicity distributions in central Au+Au collisions as a function of energy with a significance of 3.1 $ \sigma$. These results from BES-I inspired a BES-II program which focuses on the collision energy region between \mbox{3 -- 20 GeV ($750 > \mu_{\rm B} > 200$ MeV)}. BES-II combines both collider and fixed-target configurations of the STAR experiment to investigate in detail the change of behavior and understand the nature of the phase transition~\cite{STAR:2020dav}.

When studying the higher-order cumulant ratios, it is essential to demonstrate that in the absence of critical behavior, the ratios are consistent with the expectations from the non-critical baseline. The expectation for the $C_4/C_2$ ratio under Poisson statistics is unity, though the measured net-proton $C_4/C_2$ ratio within the experimental kinematic acceptance is expected to show a reduction due to the baryon number conservation~\cite{Bzdak:2012an, PhysRevC.100.034905}. This reduction is expected to increase with decreasing collision energy for fixed kinematic acceptance~\cite{Braun-Munzinger:2020jbk}. Previously, the HADES Collaboration reported a measurement of the proton $C_4/C_2$ ratio in central Au+Au collisions at \mbox{$\sqrtsNN=2.4$ GeV} consistent with unity within large uncertainties~\cite{HADES_Proton}. More data at the low collision energy is needed to quantitatively interpret the collision energy dependence of the (net-)proton fluctuation.

It was also pointed out that the experimentally measured multiplicity distributions suffer sizable contributions from fluctuating collision volume. This effect, often called volume fluctuation (VF), is due to a weak correlation between the measured reference multiplicity and the initial number of participants. It is shown in the study~\cite{Chatterjee_2021} using a hadronic transport model in $\sqrtsNN$ = 3 GeV \auau{} collisions, the centrality resolution for determining the collision centrality using charged particle multiplicities is not sufficient to reduce the initial volume fluctuation effect for higher-order cumulant analysis within current experimental acceptance. Therefore, to better understand the VF effect, it is important to systematically perform measurements within various kinematic windows and different collision centralities.

Regarding the acceptance dependence ($p_{\rm T}$ and rapidity) of cumulants and ratios, it was pointed out in Ref.~\cite{Ling:2015yau} that there may be two qualitatively different regimes: $\Delta y \gg \Delta y_{\rm corr}$ and  $\Delta y \ll \Delta y_{\rm corr}$, where $\Delta y$  is the width of the kinematic acceptance in rapidity and $\Delta y_{\rm corr}$ is the range of the proton correlations in rapidity. 
When $\Delta y \ll \Delta y_{\rm corr}$, one expects the cumulant ratios to approach the Poisson limit at $\Delta y \sim \mean{N} \rightarrow  0$. Alternatively, one expects the correlation functions to become rapidity independent as $\Delta y$ becomes wider. In the $\Delta y \gg \Delta y_{\rm corr}$ regime as $\Delta y$ increases, cumulants are expected to grow linearly for the uncorrelated contributions while the cumulant ratios are expected to saturate for any physical correlations. Therefore, the rapidity and transverse momentum dependence of proton cumulants and correlation functions are important in the search for signatures of criticality. It should be noted, that the acceptance dependence could be sensitive to non-equilibrium effects~\cite{Koch:2020azm, Bzdak:2018axe}, smearing due to diffusion and hadronic rescattering in the expansion of the system~\cite{thermal_blurring}.

In this paper, we report the cumulants and correlation functions of proton multiplicity distributions in Au+Au collisions at $\sqrtsNN$ = 3 GeV, the lowest energy of the STAR fixed-target program. The paper is organized as follows: Sec.~\ref{sec:exp} describes the experimental setup, data sets, and analysis details including corrections, systematic uncertainties, and the effect of volume fluctuation. Sec.~\ref{sec:result} presents the proton multiplicity cumulants, correlation functions, and their corresponding ratios. This includes the acceptance and energy dependence of the cumulant ratios. In addition, the data are compared to various model calculations. Finally, we summarize our findings from this analysis in Sec.~\ref{sec:summary}. 

\section{Experiment and data analysis}
\label{sec:exp}
\subsection{Data set and event selection}

The dataset analyzed in this paper was collected in 2018 by the Solenoidal Tracker at RHIC (STAR) using a fixed-target configuration. 
The gold target of the thickness of 1.93 g/cm$^2$ ($0.25$ mm) corresponding to a 1\% interaction probability was located 200.7 cm from the center of the Time Projection Chamber (TPC)~\cite{STAR:2002eio}. A beam, consisting of 12 bunches of $7\times 10^9$ gold ions is circulated in the RHIC ring at a frequency of 1 MHz with an energy of 3.85 GeV per nucleon. 

Proton multiplicities were recorded in the TPC and Time-of-Flight detectors (TOF)~\cite{LLOPE2012S110}, which are located inside STAR's solenoidal magnet. The magnet provides a uniform 0.5 T field along the beam axis.
A total of $1.4\times 10^{8}$ \auau{} events at \mbox{$\sqrtsNN$ = 3 GeV} were used in this analysis.
The minimum bias events required a hit in either the Beam-Beam Counter (BBC)~\cite{Bieser:2002ah} or the Event Plane Detector (EPD)~\cite{Adams:2019fpo} and at least three hits in the TOF.   
To remove collisions between the beam and the beam pipe, event vertices are required to be less than \mbox{1.3 cm} from the Au target along the beamline and less than \mbox{1.5 cm} from the target radially from the mean collision vertex from the TPC center along the beam line.
Events are also checked on the average of variables for different run periods: charged particle multiplicity, vertex position, and track's pseudo-rapidity $\eta$ (\mbox{$\eta = 0.5*\ln[\frac{p+p_{\rm z}}{p-p_{\rm z}}]$}, where $p$ and $p_{\rm z}$ are total momentum and its fraction in beam direction), the distance of closest approach (DCA), and transverse momentum ($p_{\rm T}$). The outlier runs which deviate more than \mbox{$\pm$3 $\sigma$} are excluded in the analysis where $\sigma$ is the standard deviation of run-by-run distributions of variables listed above. 

\begin{figure}
	\centering
	\includegraphics[width=0.45\textwidth]{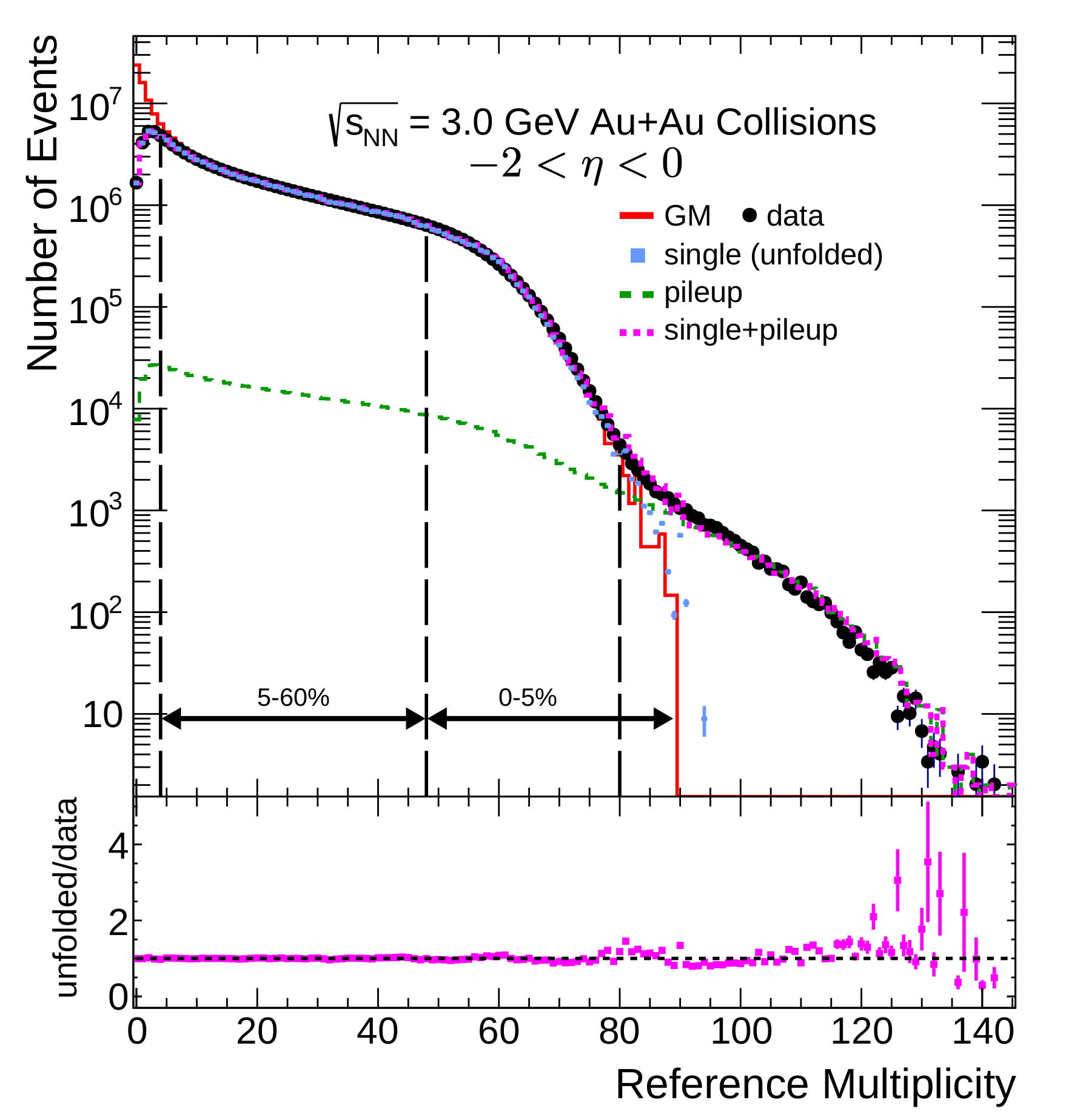}
	\caption{Reference multiplicity distributions obtained from Au+Au collisions at \mbox{$\sqrtsNN$ = 3 GeV} data (black markers), Glauber model (red histogram), and unfolding approach to separate single and pileup contributions. Vertical lines represent statistical uncertainties. Single, pileup and single+pileup collisions are shown in solid blue markers, dashed green, and dashed pink lines, respectively. The 0-5\% central events and 5-60\% mid-central to peripheral events are indicated by black arrows. The ratio of the single + pileup to the measured multiplicity distribution is shown in the lower panel.  }
	\label{fig:mult}
\end{figure}

The Au+Au collisions are characterized by their centrality. Here, the centrality is a measure of the geometric overlap of the two colliding nuclei and can be determined by measuring charged particle multiplicity in the TPC. To maximize the centrality resolution and minimize the self-correlation effect~\cite{Chatterjee:2019fey,Adam:2020unf, Abdallah:2021fzj,Chatterjee_2021}, the reference multiplicity includes charged particles except protons in the full TPC acceptance (TPC covers the pseudo-rapidity $\eta$ of \mbox{$-2<\eta<0$} in lab frame). Protons \mbox{3~$\sigma$} away from theoretical expectation are excluded by a TPC particle identification cut. The anti-proton production is negligible at \mbox{$\sqrtsNN$ = 3 GeV} and does not affect the centrality determination ($\overline{p}/p \sim \exp({-2\mu_{\rm B}/T_{\rm ch}})< 10^{-6}$)~\cite{Andronic:2017pug}. The reference multiplicity distribution in Fig.~\ref{fig:mult} is fitted with a Monte Carlo Glauber model (GM)~\cite{Miller:2007ri} coupled with a two-component model~\cite{Kharzeev:2000ph}. The two-component model assumes multiplicity $n$ in nuclear collisions has two components which are respectively proportional to the number of participants $N_{\rm part}$ and the number of binary collisions $N_{\rm coll}$:
\begin{equation}
\frac{{\rm d}n}{{\rm d}\eta} = (1-x)n_{\rm pp}\frac{\langle N_{\rm part}\rangle}{2} + x n_{\rm pp}\langle N_{\rm coll}\rangle,
\end{equation} where $x$ and $n_{\rm pp}$ denote the fraction of multiplicity from $N_{\rm coll}$ and the mean multiplicity measured in $pp$ collisions per unit of pseudo-rapidity due to $N_{\rm coll}$, respectively. Then the simulated multiplicity per event is obtained to sample $n_{\rm pp}$ times of the negative binomial distribution (shown in Eq.~\ref{eq:NBD}, where $\mu$ is the mean, and is set to $n_{\rm pp}$). 
\begin{equation}\label{eq:NBD}
P(x;\mu, k) = \frac{\Gamma(x+k)}{\Gamma(x-1)\Gamma(k)}\left(\frac{\mu/k}{1+\mu/k}\right)^{x}\frac{1}{(1+\mu/k)^k}
\end{equation}
The fit is performed by minimizing a $\chi^2$ between the measured multiplicity and the GM from the reference multiplicity from 20 to 80. The parameters for the best fit are: $n_{\rm pp}$ = 0.62, $x$ = 0.06, and $k$ = 5.56.  At reference multiplicities below $10$, the data and the GM disagree due to peripheral event trigger inefficiency. At multiplicities above $80$, double collision (pileup) events dominate the multiplicity distribution. The collision centrality is determined by fitting the Glauber calculation of charged particle multiplicity distribution to that of data. According to the normalized distribution from the Glauber model, one can extract the collision parameters such as $\langle N_{\rm part}\rangle$ and the fraction of the collision centrality, 0-5\%, 5-10\%, ..., 50-60\%. In addition to a pileup correction discussed in Sec.~\ref{sec:pileup}, events above the reference multiplicity of 80 are removed from the 0--5\% centrality class. The selection cuts for each centrality class, $\langle N_{\rm part}\rangle$ as well as the pileup fraction are shown in Tab.~\ref{tab:sys_plot}.    

\begin{table}[h!]
	\center
	\begin{tabular}{c|c|c|c} 
		\toprule
		Centrality (\%) & ~$N_{\rm ch}\geq$~ &~$\mean{N_{\rm part}}$~ &~pileup~(\%)\\
		\midrule
		0--5 & 48 & $326~(11)$   & 2.10 \\ 
		5--10 & 38 & $282~(8)$  & 1.47 \\
		10--20 & 26 & $219~(8)$  & 1.28\\
		20--30 & 16 & $157~(7)$  & 1.07 \\
		30--40 & 10 & $107~(5)$  & 0.90\\ 
		40--50 & 6 & $70~(5)$  &0.75 \\
		50--60 & 4 & $47~(5)$ & 0.64\\
		\bottomrule
	\end{tabular}
	\caption{The uncorrected number of charged particles except protons (${N_{\rm ch}}$) within the pseudo-rapidity $-2<\eta<0$ used for the centrality selection for Au+Au collisions at $\sqrtsNN$ = 3 GeV. The centrality classes are expressed in \% of the total cross-section. The lower boundary of the particle multiplicity (${N_{\rm ch}}$) is included for each centrality class. Values are provided for the average number of participants ($\mean{N_{\rm part}}$) and pileup fraction. The fraction of pileup for each centrality bin is also shown in the last column. The averaged pileup fraction from the minimum biased collisions is determined to be 0.46\%. Values in the brackets are systematic uncertainty. }
	\label{tab:sys_plot}
\end{table}

\subsection{Track selection, particle identification and acceptance}
\begin{figure*}[!htbp]
	\centering
	\includegraphics[width=1.0\textwidth]{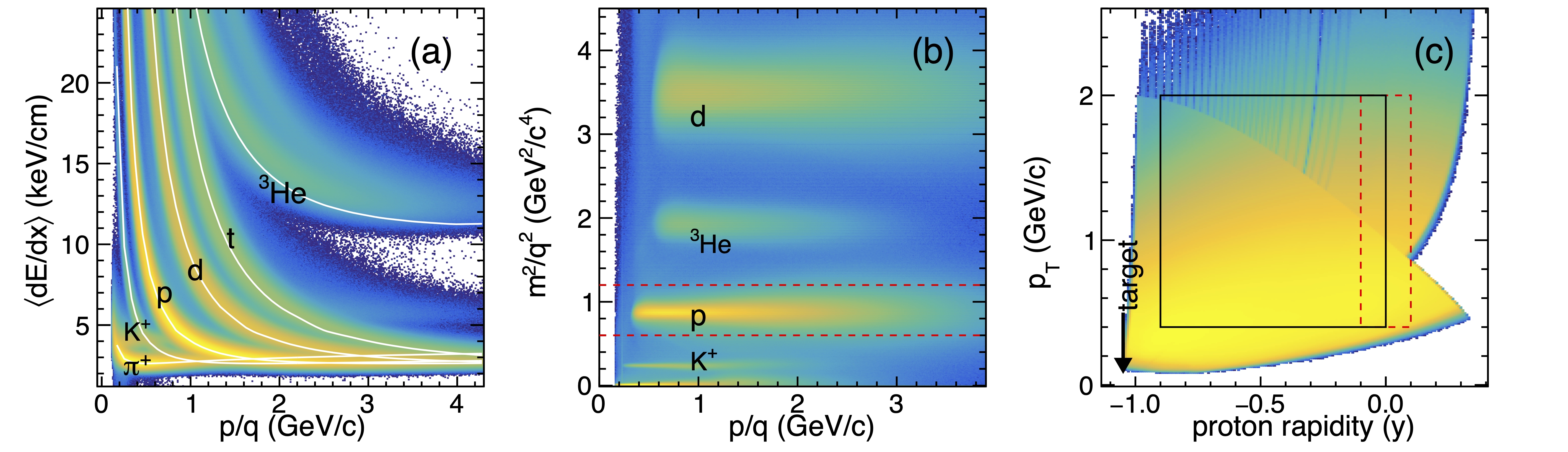}
	\caption{(a):  $\langle {\rm d}E/{\rm d}x\rangle$ vs. particle rigidity measured in the TPC; pion, kaon, proton and deuteron bands are labeled. The proton is plotted in red from the Bichsel formula. (b): Mass-squared vs. the particle rigidity measured in the TPC and TOF. Kaon, proton, deuteron, and Helium-3 peaks are labeled. The red dashed lines indicate selection cuts by mass-squared. (c): Analysis acceptance in transverse momentum vs. proton rapidity ($y$) in the center-of-mass frame Au+Au collisions at $\sqrtsNN$ = 3 GeV. Black box indicates acceptance for rapidity $-0.9 < y < 0$ and momentum $0.4<p_{\rm T}<2.0$ GeV/$c$. The Red dashed box indicates a narrower rapidity window $|y| < 0.1$, the largest possible symmetric rapidity window from this data set.}
	\label{fig:acceptance_pid}
\end{figure*}
The TPC measures both the trajectory and the energy loss (${\rm d}E/{\rm d}x$) of a particle. 
TPC spatial hits are fitted with helices to determine the charge and momentum of each charged particle. To ensure track quality, tracks are required to meet selection criteria which are at least 10 hits and more than five ${\rm d}E/{\rm d}x$ measurements.
Additionally, to prevent double-counting reconstructed tracks from a single particle, a selected track is required to have more than 52\% of the maximum-possible fit points, which peaks at 45 possible hits.  
To suppress the contamination from spallation in the beam pipe and secondary protons from hyperon decays, DCA $<$ 3~cm criterion is placed at the distance of the closest approach (DCA) in 3-dimensions of the reconstructed track's trajectory to the primary vertex position. The results presented here are within the kinematics $-0.9<y<0$ and $0.4<\ppt<2.0$ GeV/$c$.

Particle identification (PID) is performed by measuring the ${\rm d}E/{\rm d}x$ and the time of flight in the TPC and TOF, respectively. Figure \ref{fig:acceptance_pid} (a) shows the $\langle {\rm d}E/{\rm d}x\rangle$ as a function of rigidity (the ratio of total momentum over electric charge, $|p|/q$ GeV/$c$) for the positively charged tracks. To select proton candidates, the measured values of ${\rm d}E/{\rm d}x$ are compared to a theoretical prediction~\cite{BICHSEL2006154} (red line). The quantity $N_{\sigma,p}$ for charged tracks in the TPC is defined as 
\begin{equation}\label{eq:tpc_nsigma}
	N_{\sigma,p} = \frac{1}{\sigma_{\rm R}}\ln{\frac{\mean{{\rm d}E/{\rm d}x}}{\mean{{\rm d}E/{\rm d}x}^{\rm th}}},
\end{equation}
where $\mean{{\rm d}E/{\rm d}x}$ is the truncated mean value of the track energy loss measured in the TPC, $\mean{{\rm d}E/{\rm d}x}^{\rm th}$ is corresponding theoretical prediction, and $\sigma_{\rm R}$ is the track length dependent ${\rm d}E/{\rm d}x$ resolution. The $N_{\sigma,p}$ distribution appears as a standard Gaussian distribution with a mean close to zero. The offset from zero is measured as a function of momentum in 0.1 GeV/$c$ bins and the $N_{\sigma,p}$ distribution is re-centered. The proton tracks are selected within three standard deviations of the re-centered $N_{\sigma,p}$ distribution ($|N_{\sigma,p}|<3.0$).  

Figure~\ref{fig:acceptance_pid}~(b) shows the mass-squared ($m^2$) versus rigidity of charged particles in the TPC and TOF. The $m^2$ is given by 
\begin{equation}
	m^2 = p^2 \left( \frac{c^2t^2}{L^2} - 1 \right),
\end{equation}
where $p$, $t$, and $L$ are the momentum, time of flight, and path length of the particle, respectively. The speed of light in vacuum is denoted by $c$. The protons are identified by selecting charged tracks with mass-squared values between $0.6<m^2<1.2$ GeV$^2$/$c^4$. 
While the mass-squared cut provides high proton purity, it introduces a 60\% matching efficiency. 
The proton purity is required to be higher than 95\% at all rapidities and momenta for the subsequent cumulant analysis.

Figure~\ref{fig:acceptance_pid}~(c) shows the transverse-momentum versus rapidity for protons selected in the TPC within $|N_{\sigma,p}|<3.0$. The tracks above a momentum of 2.0 GeV/$c$ in the lab frame are required to have a mass-squared cut.
The kinematic acceptance of the analysis ($-0.9<y<0$ and $0.4<\ppt<2.0$ GeV/$c$) is indicated by a black box. 

\begin{figure*}%[!htbp]
	\centering
	\includegraphics[width=0.9\textwidth]{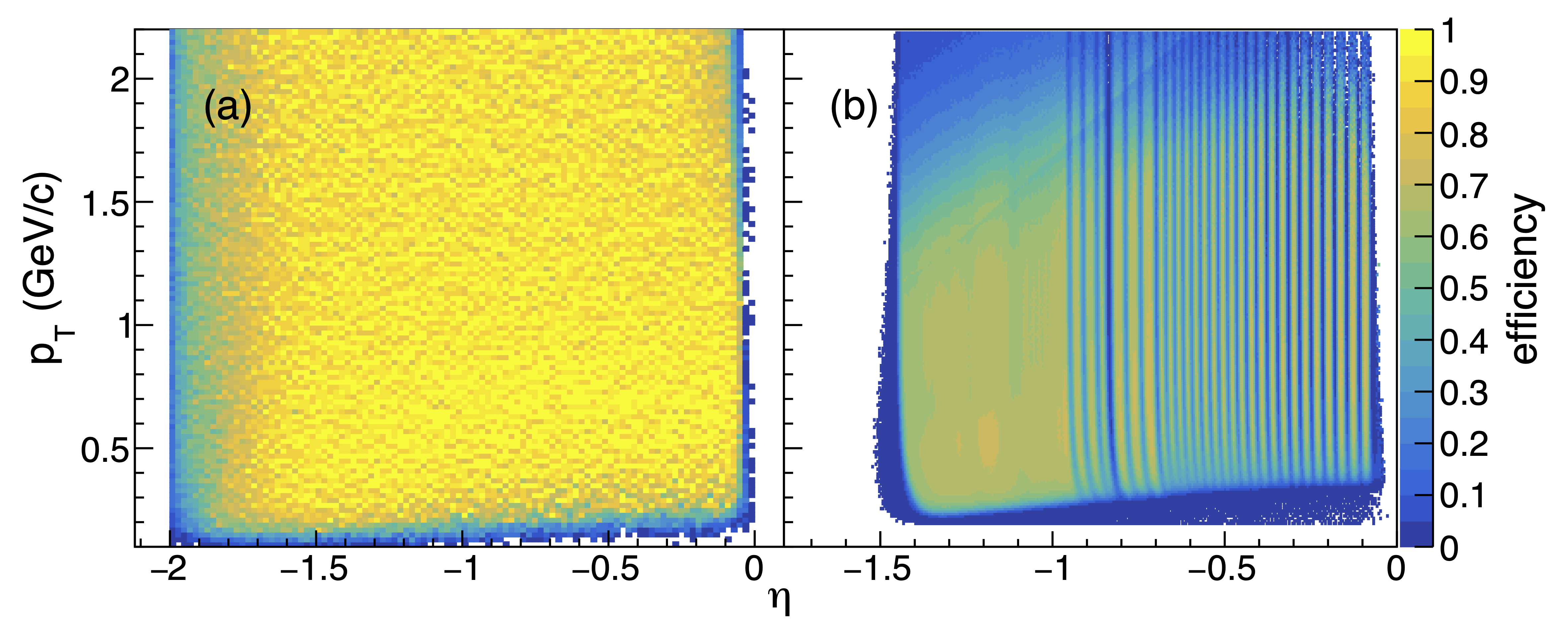}
	\caption{ (a): TPC detector efficiency for protons measured from simulation in $\ppt$ vs. pseudo-rapidity. The shared z-axis indicates track efficiency. (b): TOF detector matching efficiency for protons measured in $\ppt$ vs. pseudo-rapidity. }
	\label{fig:eff_map}
\end{figure*}

\subsection{Definition of cumulants and correlation functions}\label{cum:def}

Here, the definition of cumulants is provided. Let $N$ be the number of particles measured in each event. Then mean value of $N$ is given by $\mean{N}$ and $\delta N = N - \mean{N}$ is the deviation from the mean value where the $\langle .\rangle$ symbol indicates the average over events. The $r^{\rm th}$-order central moment of a distribution is described by
\begin{equation}
	\mu_{r} = \mean{(\delta N)^{r}}\quad {r}\geq2.
\end{equation} 
In terms of central moments, the cumulants are defined as 
\begin{equation}\label{eq:cumulant1}
	\begin{split}
		C_1 &= \mean{N}, \\
		C_2 &= \mean{(\delta N)^2} = \mu_2, \\
		C_3 &= \mean{(\delta N)^3} = \mu_3, \\
		C_4 &= \mean{(\delta N)^4} - 3\mean{(\delta N)^2}^2 = \mu_4 - 3\mu_2^2,\\
		C_5 &= \mean{(\delta N)^5} - 10\mean{(\delta N)^2}\mean{(\delta N)^3}\\&=\mu_5-10\mu_2\mu_3,\\
		C_6 &= \mean{(\delta N)^6}+30\mean{(\delta N)^2}^3\\&-15\mean{(\delta N)^2}\mean{(\delta N)^4}-10\mean{(\delta N)^3}^2\\&=\mu_6+30\mu_2^{3}-15\mu_2\mu_4-10\mu_3^{2},\\
C_{n} &= \mu_{n} - \sum^{n-2}_{m=2}\binom{n-1}{m-1}C_{m}\mu_{n-m},\quad {n}>3.
	\end{split}
\end{equation} The cumulants can also be expressed in terms of raw moments (Eq.~\ref{eq:a8} in Appendix~\ref{app:a}).  
Some commonly used moments and ratios are given as 
\begin{equation}
	\begin{split}
		M &= C_1, \\
		S &= \frac{C_3}{(C_2)^{3/2}},
	\end{split}
	\quad
	\begin{split}
		\sigma^2 &= C_2, \\ 
		\kappa &= \frac{C_4}{C_2^2},
	\end{split}
\end{equation} where $M$, $\sigma^2$, $S$, and $\kappa$ are mean, variance, skewness, and kurtosis, respectively.
The products $S\sigma$ and $\kappa\sigma^2$ can be expressed in terms of the cumulant ratios as
\begin{equation}
	\sigma^2/M = \frac{C_2}{C_1}, \quad S\sigma = \frac{C_3}{C_2}, \quad \kappa\sigma^2 = \frac{C_4}{C_2}.
\end{equation}

In case there are no intrinsic correlations among the measured particles, all ratios of the cumulants are unity, thus Poisson statistics is a non-trivial baseline for experimentally measured cumulant ratios.

The probability distribution of Poisson statistics is $P(N)=\lambda^N e^{-\lambda}/N!$, where $\lambda$ is an average of the number of measured particles per event. The cumulants are equal to the mean value: $C_1=C_2=...=C_{n}=\lambda$, where $C_{n}$ is the $n^{\rm th}$-order cumulant. Furthermore, all cumulant ratios equal one. More discussion can be found in Ref.~\cite{Asakawa:2015ybt} and Appendix~\ref{app:b}.

As discussed in Refs.~\cite{Bzdak:2016jxo,Abdallah:2021fzj}, the cumulants $C_{r}$ can be algebraically converted to the integrals of the corresponding multi-particle correlation functions. These integrated correlation functions also known as factorial cumulants will be simply called correlation functions (denoted by $\kappa_{i}$) henceforth. In terms of cumulants, the correlation functions up to $6^{\rm th}$-order are
\begin{equation}
	\begin{split}
		\kappa_{1} &=C_{1},\\
		\kappa_{2} &=-C_{1}+C_{2},  \\
		\kappa_{3} &=2C_{1}-3C_{2}+C_{3},  \\
		\kappa_{4} &=-6C_{1}+11C_{2}-6C_{3}+C_{4}, \\
		\kappa_{5} &=24C_{1}-50C_{2}+35C_{3}-10C_{4}+C_{5},\\
		\kappa_{6} &=-120C_{1}+274C_{2}-225C_{3}+85C_{4}\\&-15C_{5}+C_{6}.
		\label{eq:kappa}
	\end{split}
\end{equation}
A compact form of the above equations can be seen in Eq.~\ref{eq:cfc} of Appendix~\ref{app:a}.

We define $C_{i}/{C_1}-1,\quad {i} =1, 2, \cdots$ as reduced cumulant ratio. The reduced cumulant ratio of different orders can be displayed in terms of correlation function ratios as

\begin{align}
	\begin{split}\label{eq:kappa1}
		\frac{C_2}{C_1} - 1 &= \frac{\kappa_2}{\kappa_1},\\
		\frac{C_3}{C_1} - 1 &= 3\frac{\kappa_2}{\kappa_1} + \frac{\kappa_3}{\kappa_1},\\
		\frac{C_4}{C_1} - 1 &=7\frac{\kappa_2}{\kappa_1}+6\frac{\kappa_3}{\kappa_1}+\frac{\kappa_4}{\kappa_1},\\
		\frac{C_5}{C_1} - 1 &= 15\frac{\kappa_2}{\kappa_1}+25\frac{\kappa_3}{\kappa_1}+10\frac{\kappa_4}{\kappa_1}+\frac{\kappa_5}{\kappa_1},\\
		\frac{C_6}{C_1} -1 &= 31\frac{\kappa_2}{\kappa_1}+90\frac{\kappa_3}{\kappa_1}+65\frac{\kappa_4}{\kappa_1}+15\frac{\kappa_5}{\kappa_1}+\frac{\kappa_6}{\kappa_1}.
	\end{split}
\end{align}

It is clear that the $n^{\rm th}$-order reduced cumulant ratio is a combination of all multi-particle correlation functions up to the $n^{\rm th}$-order.

\begin{figure*}
	\centering
	\includegraphics[width=1.0\textwidth]{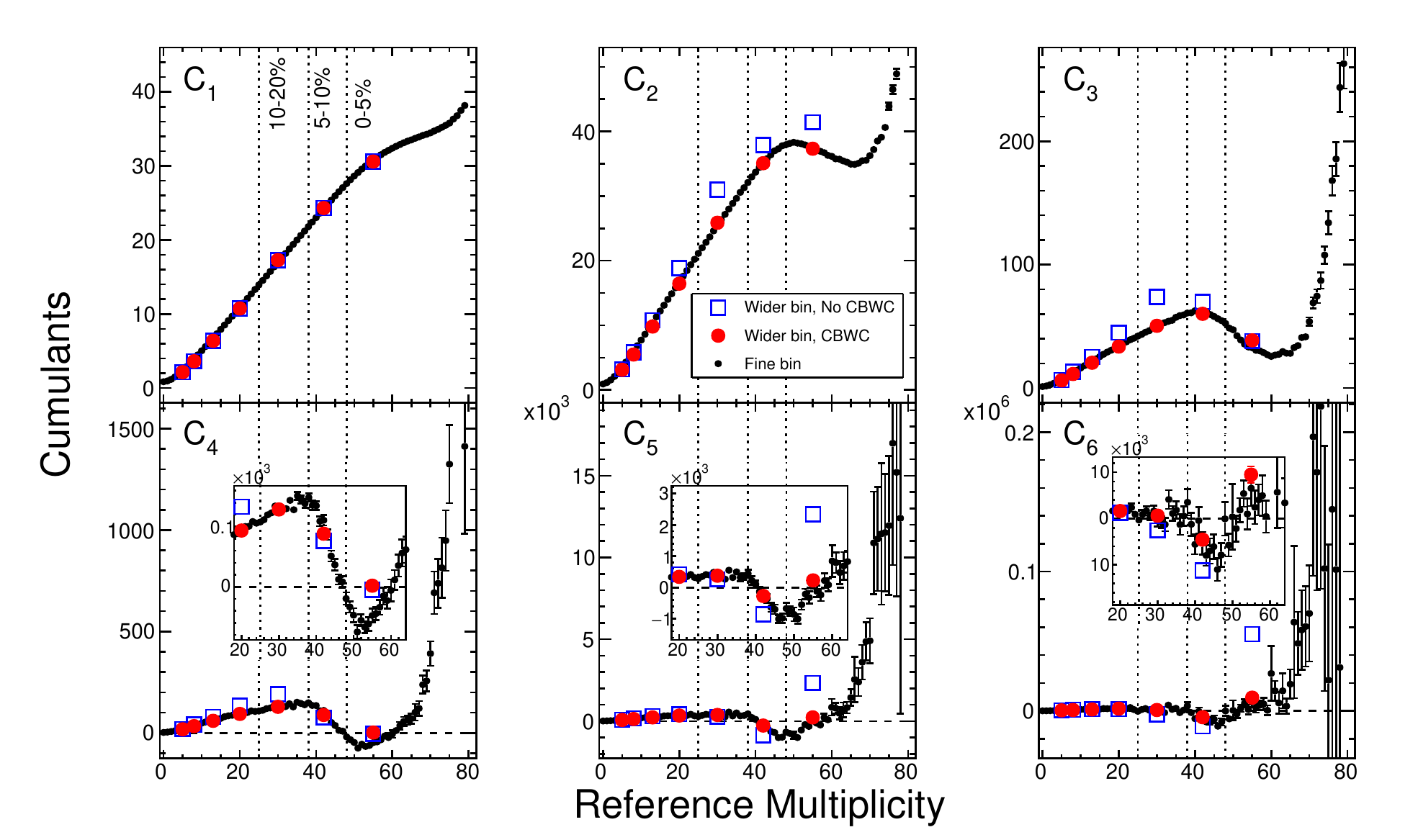}
	\caption{Proton cumulants as a function of reference multiplicity (black circles) from $\sqrtsNN$ = 3 GeV Au+Au collisions. Centrality-binned results with and without centrality bin width corrections are represented by red circles and blue squares, respectively. Vertical dashed lines indicate the centrality classes, from right to left: $0-5\%$, $5-10\%$, $10-20\%$.
	Data points in this figure are only corrected for detector efficiency but not for the pileup effect which will be discussed in the later section.}
	\label{fig:cbwc}
\end{figure*}

\subsection{Detector efficiency correction}

In this analysis, the proton tracks are corrected for detector inefficiency in the TPC and TOF. The TPC efficiency is calculated by placing Monte Carlo tracks into a GEANT~\cite{Fine:2000qx} detector simulation. The GEANT detector response is mixed with the detector response from real data and a track reconstruction process is performed. The TPC efficiency is calculated by counting the fraction of successfully reconstructed Monto Carlo tracks. Figure \ref{fig:eff_map} (a) shows the TPC detector efficiency for proton tracks with respect to kinematic acceptance ($p_{\mathrm{T}}$ versus $\eta$). 
The TOF matching efficiency is estimated directly from data by measuring the fraction of TPC tracks that satisfy the TOF-matching criteria.
Recall, a TOF-matched proton candidate requires $p > 2$ GeV/$c$, $|N_{\sigma,p}| < 3.0$, and $0.6<m^2<1.2$ GeV$^2$/$c^4$. Figure \ref{fig:eff_map} (b) shows TOF matching efficiency for proton tracks in a window of $p_{\mathrm{T}}$ versus $\eta$.
The detector efficiency corrections are performed on a ``track-by-track'' basis~\cite{PhysRevC.95.064912,Luo:2019}, where the proton reconstruction efficiency as a function of $\ppt$ and rapidity is applied as a weight to each track. 

%--======================================================================
\subsection{Centrality bin width correction}

The proton cumulants are evaluated in an event-by-event manner. To extract an averaged value of the cumulants from a range of the measured reference multiplicities, or in other words, from a centrality bin, a proper procedure called centrality bin width correction (CBWC)~\cite{Luo:2013bmi} method is applied. The number of events from each multiplicity bin is used as a weight in the averaging procedure. Figure \ref{fig:cbwc} shows centrality dependence of proton cumulants up to $6^{\rm th}$-order in Au+Au collisions at $\sqrtsNN$ = 3 GeV. The black circles show the multiplicity dependence of the cumulants, while red circles and blue squares are centrality binned cumulants with and without CBWC, respectively. The CBWC is necessary to extract properly averaged cumulants in a given centrality bin.

\subsection{Pileup correction}\label{sec:pileup}
\begin{figure}
	\centering
	\includegraphics[width=0.45\textwidth]{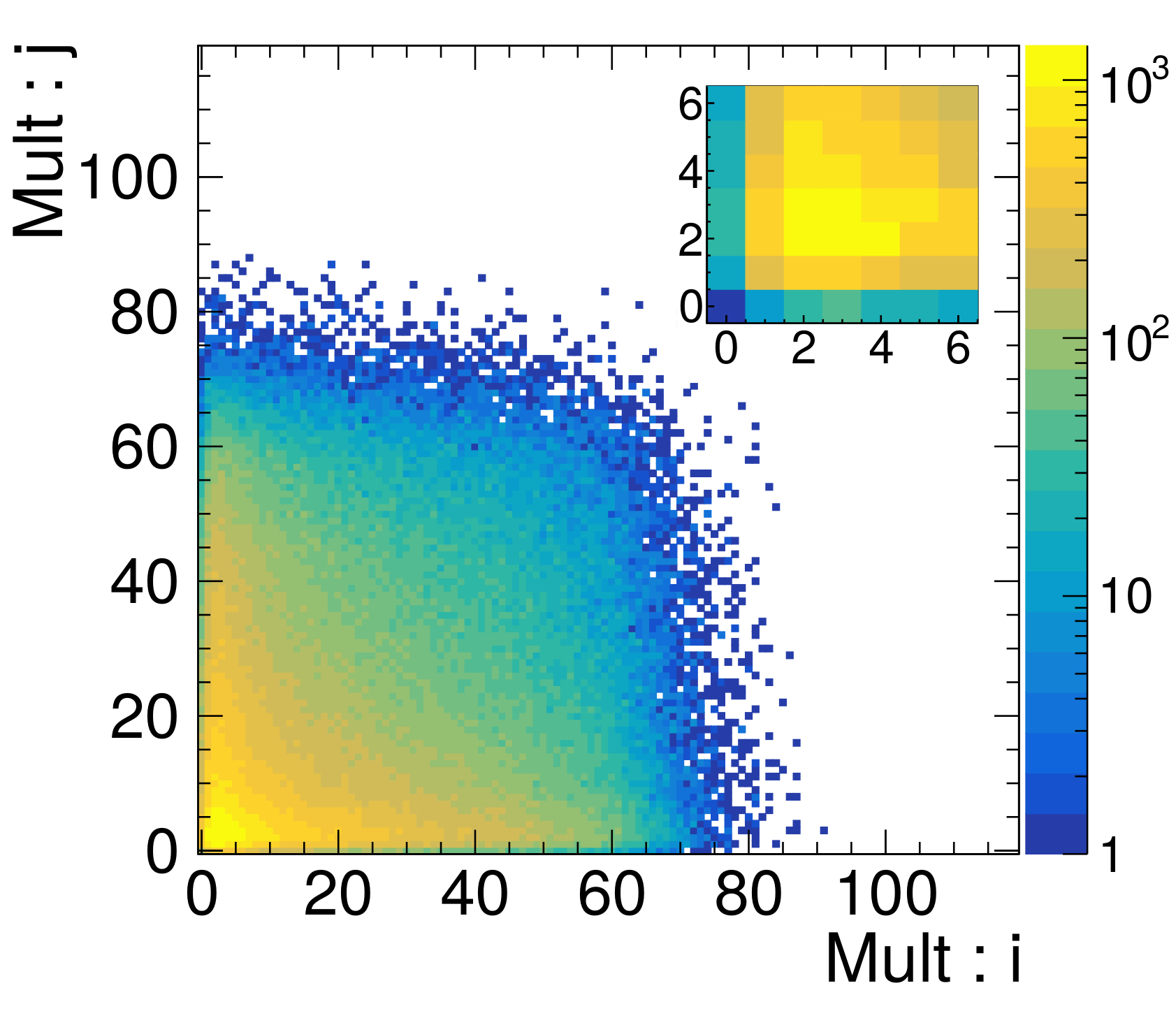}
	\caption{Correlation between reference multiplicity $i$ and $j$ from single-collision events. The small panel at the top right corner is an expanded plot with $i<7$ and $j<7$.}
	\label{fig:RM}
\end{figure}

\begin{figure*}
	\centering
	\includegraphics[width=1.0\textwidth]{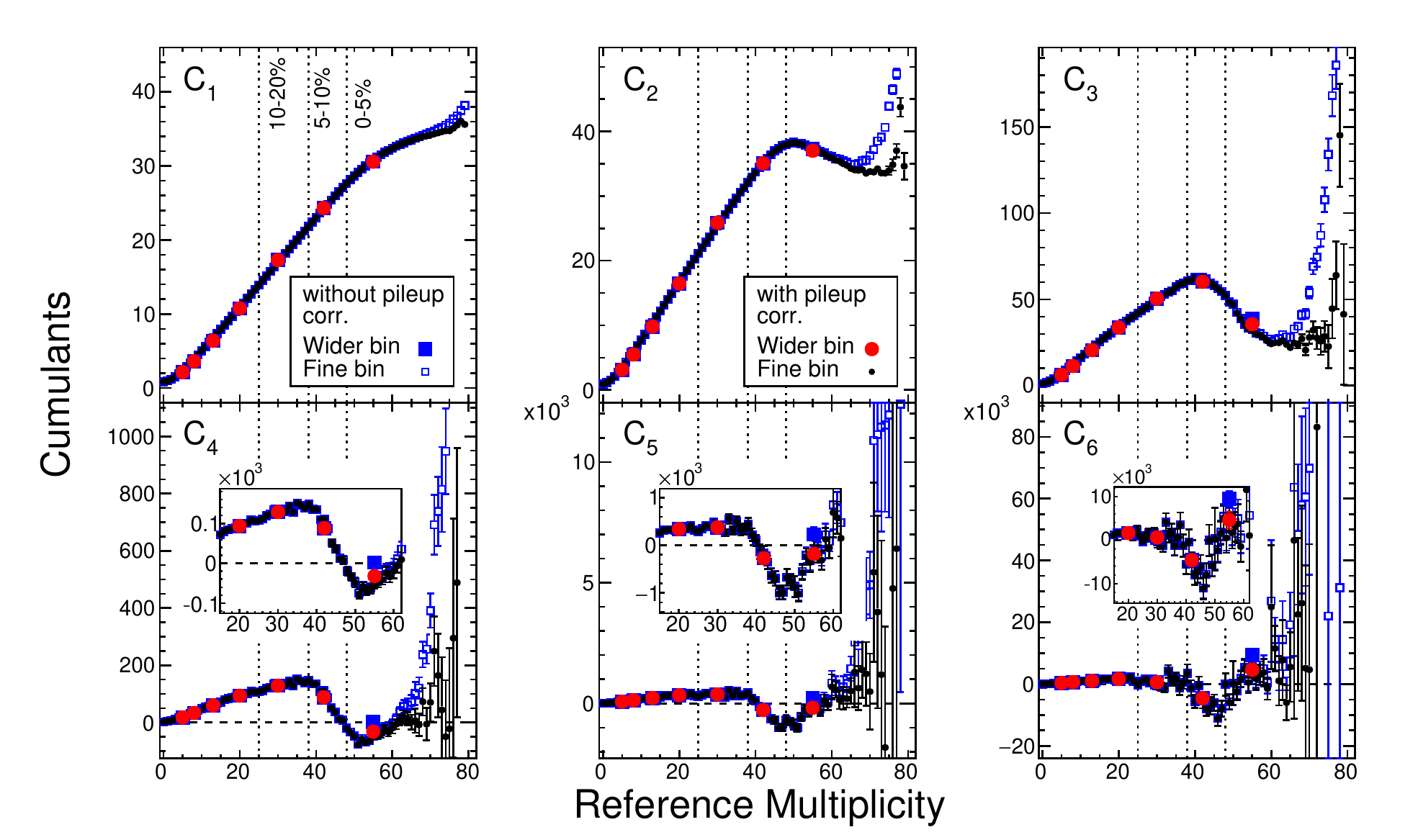}
	\caption{Proton cumulants as a function of reference multiplicity from $\sqrtsNN$ = 3 GeV Au+Au collisions. Pileup corrected and uncorrected cumulants as a function of reference multiplicity are represented by black circles and blue open squares, respectively. Red circles and blue-filled squares represent the results of centrality binned data.
}
	\label{fig:pu}
\end{figure*}

\begin{figure*}
	\centering
	\includegraphics[width=1.0\textwidth]{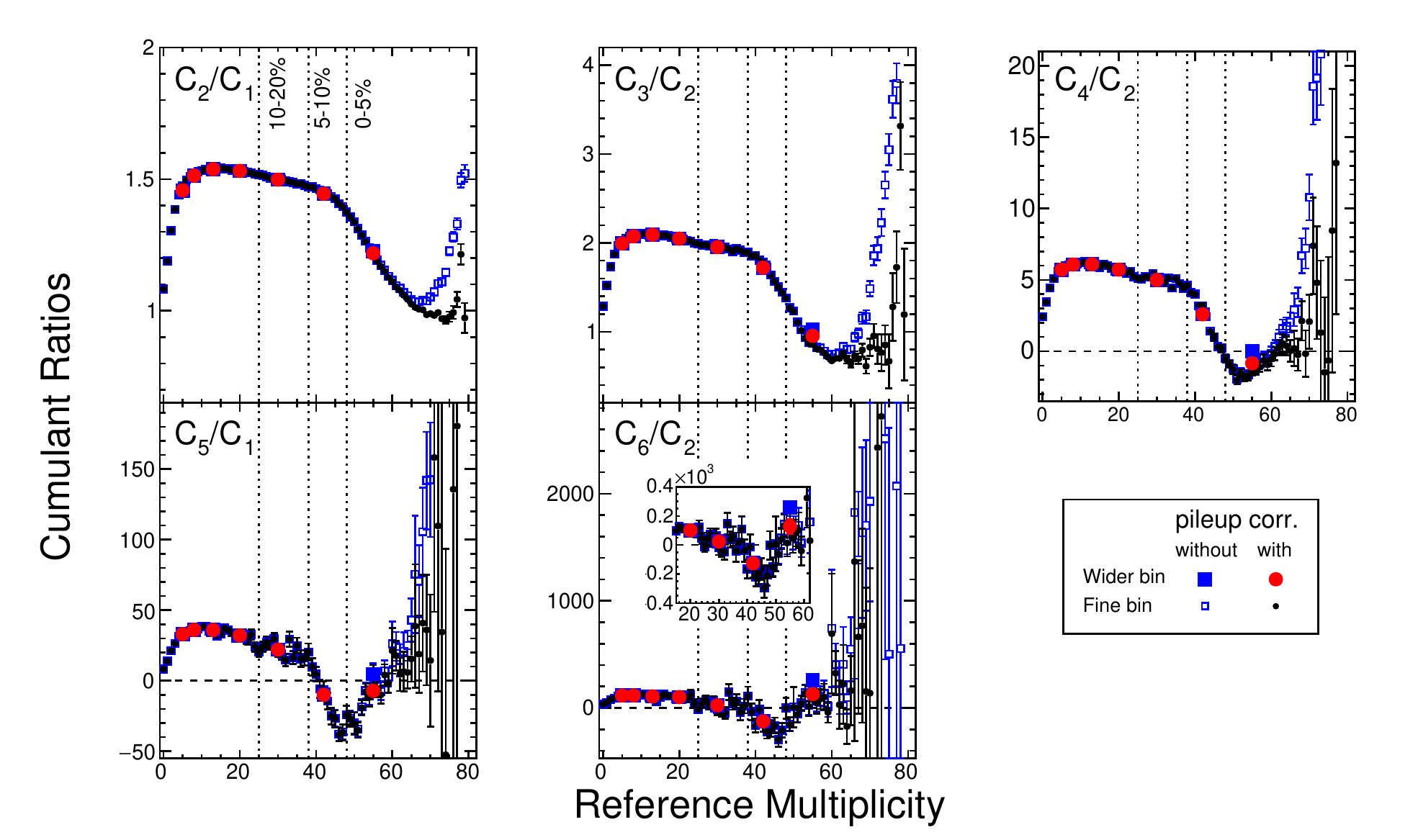}
	\caption{Ratios of proton cumulants as a function of reference multiplicity from $\sqrtsNN$ = 3 GeV Au+Au collisions. Pileup corrected and uncorrected cumulants are represented by black circles and blue open squares, respectively. Red circles and blue-filled squares represent the results of centrality binned data.
}
	\label{fig:pu_ratios}
\end{figure*}
Double collisions (``pileup''), seen in Fig.~\ref{fig:mult}, are the largest source of background in the STAR fixed-target experiment. Here, we discuss the correction technique~\cite{Nonaka:2020qcm} used to remove the pileup statistically. The correction technique requires an estimate of the pileup contribution as a function of reference multiplicity. Thus, an unfolding method~\cite{Zhang:2021rmu} used to estimate the event-averaged pileup fraction is discussed. 

The correction method assumes a pileup event is the superposition of two independent single-collision events. Let $P_{m}(N)$ be the probability distribution function to find an event with $N$ particles at multiplicity $m$.
If the probability of a pileup event at the $m^{\rm th}$ multiplicity bin is $\alpha_{m}$, then $P_{m}(N)$ is  
\begin{equation}
	P_{m}(N) = (1 - \alpha_{m})P^{t}_{m}(N) + \alpha_{m} P^{pu}_{m}(N),
\end{equation}    
where $P^{t}_{m}(N)$ and $P^{pu}_{m}(N)$ are the single-collision and pileup probability distribution functions, respectively.

The pileup events can be decomposed into the sub-pileup event probability distribution function $P^{sub}_{i,j}(N)$ as 
\begin{equation}
	P^{pu}_{m}(N) = \sum_{i,j}\delta_{m,i+j} w_{i,j} P^{sub}_{i,j}(N),
\end{equation}
where $w_{i,j}$ is the probability to observe the sub-events among all pileup events at multiplicity $m$, where $m=i+j$. Additionally, the sum over $i$ and $j$ runs over non-negative integers and $\displaystyle\sum_{i,j} \delta_{m,i+j} w_{i,j} = 1$, where $w_{i,j} = w_{j,i}$.

Following the procedure outlined in Ref.~\cite{Nonaka:2020qcm}, the single collision moments can be recursively expressed in terms of the measured moments of lower multiplicity bins as 
\begin{equation} \label{true_moments}
	\mean{N^{r}}^{t}_{m} = \frac{ \mean{N^{r}}_{m} - \alpha_{m} \beta^{({r})}_m }{ 1 - \alpha_{m} + 2 \alpha_{m} w_{{m},0}},
\end{equation}

where $\beta^{({r})}_{m}$ is defined as
\begin{equation} \label{eq:pu_cumu_mu_mom}
	\beta^{({r})}_{m} = \mu^{({r})}_{m} + \sum_{i,j>0}\delta_{m,i+j} w_{i,j} \mean{N^{r}}^{sub}_{i,j},
\end{equation}
and 
\begin{align} \label{eq:pu_mom_gen}
	\mu^{(r)}_{m} = 
	\begin{cases}
		2 w_{{m},0} \displaystyle\sum^{r-1}_{k=0} \binom{r}{k} \mean{N^{r-k}}^t_0 \mean{N^{k}}^{t_m} &({m}>0)  \\[1ex]
		\displaystyle\sum^{r-1}_{k=1} \binom{r}{k} \mean{N^{r-k}}^t_0 \mean{N^{k}}^{t}_0           &(m=0).
	\end{cases}
\end{align}

The correction requires both $\alpha_{m}$ and $w_{i,j}$ to be determined with a high level of precision. Both parameters can be expressed in terms of the multiplicity of the single collision events $T(m)$ as
\begin{align}
	w_{i,j} =& \frac{\alpha T({i})T({j})}{\sum_{i,j} \delta_{m,i+j} \alpha T({i})T({j})},\\
	\alpha_{m} =& \frac{\alpha\sum_{i,j} \delta_{m,i+j}T({i})T({j})}{(1-\alpha)T({m}) + \alpha\sum_{i,j} \delta_{m,i+j}T({i})T({j})},
\end{align}   
where $\alpha$ is the total pileup fraction overall reference multiplicities.
Therefore, the accuracy of $\alpha_{m}$ and $w_{i,j}$ is determined by one's ability to extract the single collision distribution from the measured reference multiplicity.

For this analysis, an unfolding technique~\cite{Esumi:2020xdo} is used to estimate $T({m})$. An overview of the unfolding procedure and a closure test of simulated events can be found in Ref.~\cite{Zhang:2021rmu}. The unfolding is performed by generating both a pileup distribution and single collision distribution from Monte-Carlo (toy-MC) events. The difference between the toy-MC (single + pileup) distribution and the data multiplicity distribution is measured and propagated back to the toy-MC single collisions. The process is repeated until the toy-MC and data agree. The bottom panel of Fig.~\ref{fig:mult} shows the ratio of the data and toy-MC after 100 iterations. In the top panel of Fig.~\ref{fig:mult}, the single collision and pileup distributions are represented by blue and green dashed lines, respectively. The procedure has one free parameter, which is the total pileup probability $\alpha$ in Eq.~\ref{eq:pu_mom_gen}. The procedure is run for various $\alpha$ parameters and a $\chi^2$ test is performed. The pileup probability $\alpha$ is determined to be ($0.46 \pm 0.09$)\% for all events and ($2.10 \pm 0.40$)\% in the 0--5\% centrality class. With the unfolded single collision distribution and the $\alpha$ parameter, the response matrix $w_{i,j}$ can be simulated as shown in Fig.~\ref{fig:RM}. As stated, $w_{i,j}$ is the probability to observe a sub-pileup event at multiplicity $m$ with $m=i+j$. The pileup corrected cumulants are shown in Fig.~\ref{fig:pu}. 
Additionally, the event-averaged pileup corrected (red) and uncorrected (blue) cumulants are displayed. For all cumulants, only results from the top centrality class (0-5\%) are affected. Figure \ref{fig:pu_ratios} are the pileup corrected and uncorrected cumulant ratios. Similar to the cumulants, the cumulant ratios are only affected in the most central collisions. Pileup correction will increase uncertainties in the high multiplicity region, especially for reference multiplicity larger than 60. After the pileup correction, higher-order cumulant ratios, $C_{4}/C_{2}$, $C_{5}/C_{1}$ and $C_{6}/C_{2}$, are consistent with zero within uncertainty for the most central multiplicity bins.

%--=======================================================================
\subsection{Effects of volume fluctuation}
\begin{figure}[!htbp]
	\centering
	\includegraphics[width=0.5\textwidth]{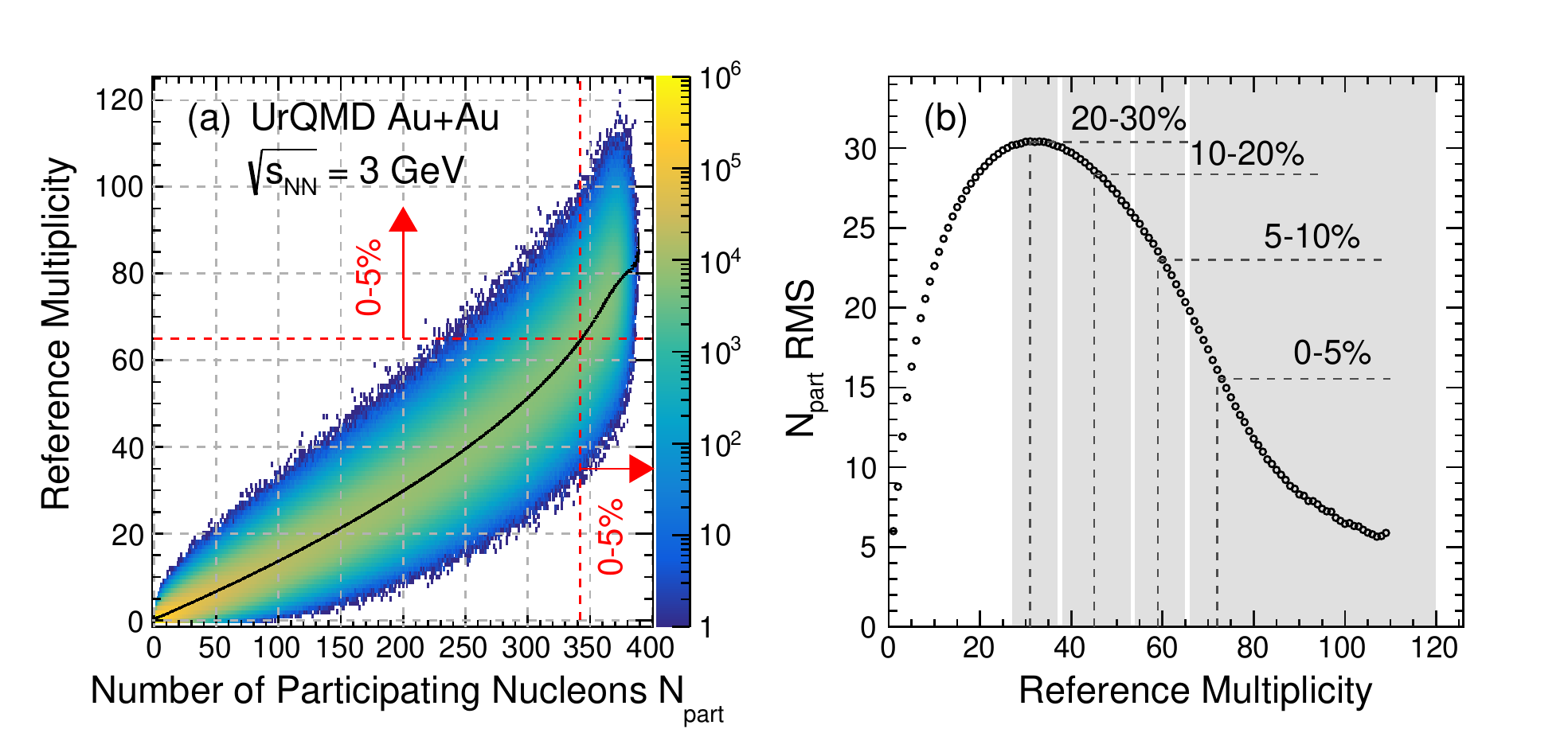}
	\caption{(a): Correlation distribution of ${N}_{\rm part}$ vs. reference multiplicity from UrQMD model. Vertical and horizontal dashed lines indicate the 0-5\% central collisions selected by ${N}_{\rm part}$ and reference multiplicity, respectively.  (b): $N_{\rm part}$ root-mean-square (RMS) distribution as a function of the reference multiplicity. The vertical lines indicate the average reference multiplicity for each centrality class.}
	\label{fig:urqmd_npart_width}
\end{figure}

\begin{figure*}[!htbp]
	\centering
	\includegraphics[width=0.6\textwidth]{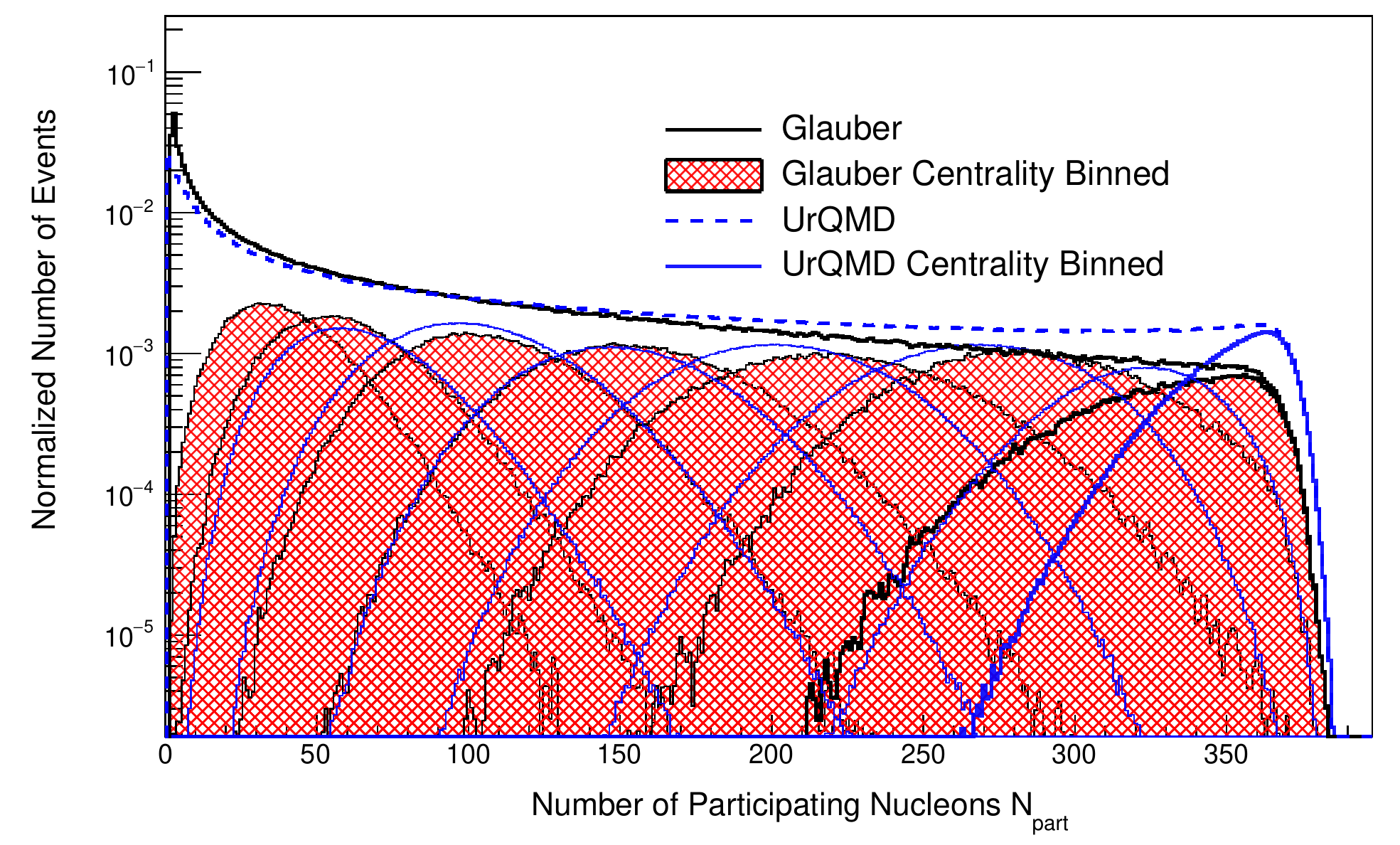}
	\caption{ $N_{\rm part}$ distributions from Monte Carlo Glauber and UrQMD model calculations. The red shaded areas and solid blue lines represent $N_{\rm part}$ distributions for $0-5\%$, $5-10\%$, $10-20\%$, $20-30\%$, $30-40\%$, $40-50\%$ and $50-60\%$ centrality classes determined by reference multiplicity.
	}
	\label{fig:gm_urqmd_npart}
\end{figure*}

\begin{figure*}
	\centering
	\includegraphics[width=0.9\textwidth]{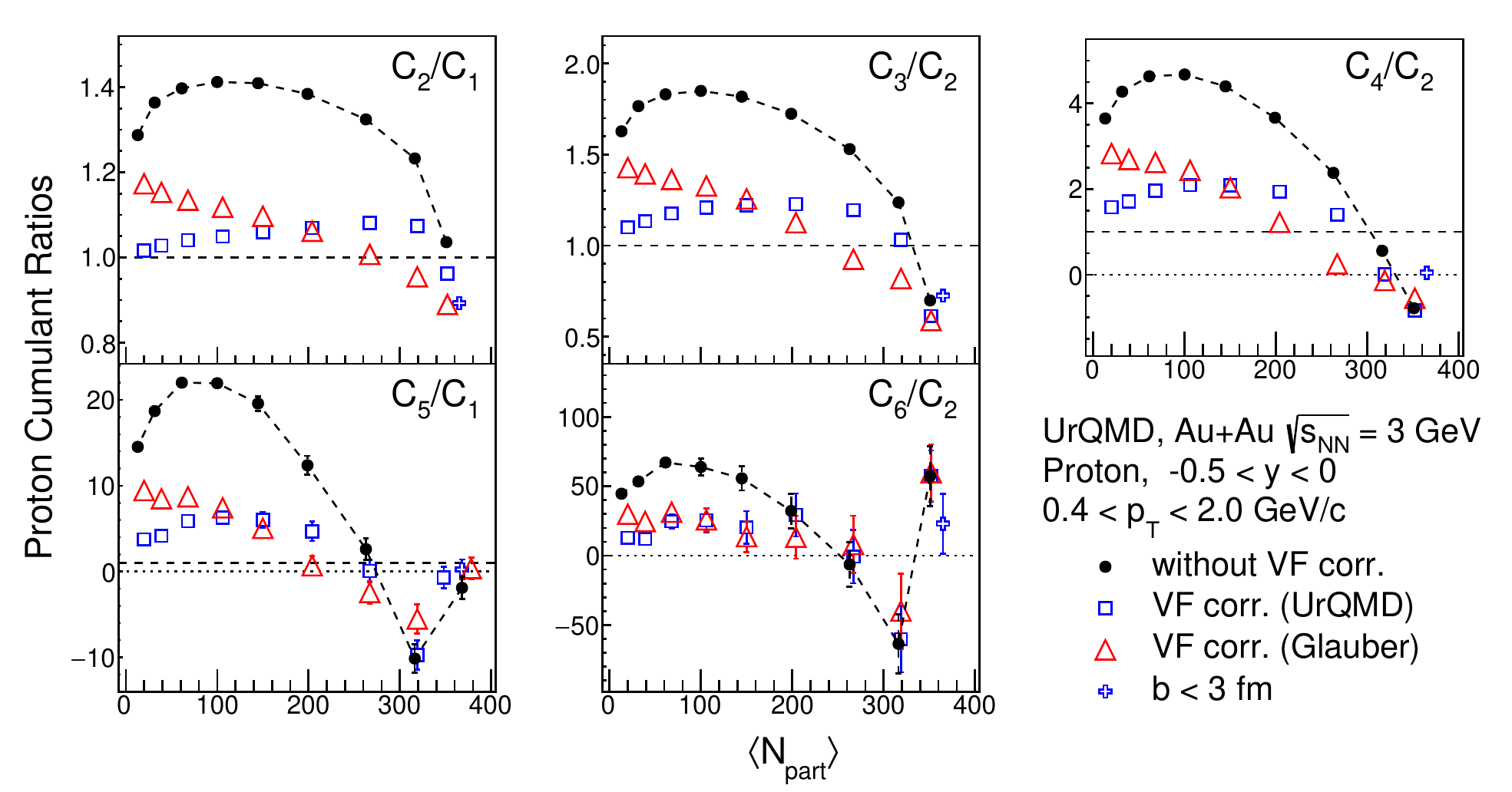}
	\caption{UrQMD results of the proton cumulant ratios up to $6^{\rm th}$-order in Au+Au collisions at $\sqrtsNN$ = 3 GeV. The black circles are without VF correction while blue squares and red triangles are results with VFC which used $N_{\rm part}$ distributions from UrQMD and Glauber model, respectively. The blue crosses are calculations using UrQMD events with $b\le3\ \rm fm$. The above results are applied CBWC except for the one (blue crosses) using $b\le3\ \rm fm$ events.}
	\label{fig:urqmd_vfc_ratio}
\end{figure*}
Physics results will be discussed for a given event centrality class. Since the physics of higher-order cumulants and their ratios are supposed to be sensitive to collision dynamics including the centrality, it is important to understand the correlation between the experimentally measured reference multiplicity distribution and the extracted class of collision centrality.
It is well-known that quantum fluctuations in particle production and fluctuation of the participating nucleon pairs will affect the final centrality determination, especially at low-energy collisions. 
The microscopic hadronic transport model UrQMD (v3.4)~\cite{Bass:1998ca, Bleicher:1999xi}, which does not contain critical phenomena physics, has been used to show the volume fluctuation effect. 
As an illustration, the UrQMD model results on the correlation of the reference multiplicity and participating nucleons $N_{\rm part}$ is shown in the left panel of Fig.~\ref{fig:urqmd_npart_width}. The right panel shows the root-mean-square (RMS) values of the $N_{\rm part}$ distribution at a given fixed reference multiplicity.

As one can see, the correlation is broad and the dispersion (RMS) of $N_{\rm part}$ is as large as 30 in the mid-central collisions for 3 GeV Au+Au collisions. Even in the most central 5\% collisions, the dispersion is in the range of 15. Primarily, the large dispersion is due to the fluctuation of the $N_{\rm part}$ in addition to the variation in the charged particle production for a given pair of nucleons. The variation of the initial number of participants for fixed reference multiplicity is also called initial volume fluctuation (IVF) and its implications on the results of the higher moments of proton distributions will be discussed later in the paper. In fact, as indicated by the red dashed lines in the plot(a), the top 5\% central collisions are largely different events with a small overlap.  

The $N_{\rm part}$ 
distributions are not measurable experimentally but obtained from the calculations of the Glauber and UrQMD models. Figure~\ref{fig:gm_urqmd_npart} shows the distributions from both Glauber (black solid line and red shaded area) and UrQMD (blue solid line and blue dashed lines) models. The hatched areas are corresponding to various collision centralities determined from the charged particle reference multiplicity. It is obvious that the overall distributions are quite different. More so, the widths of the top 5\% are dramatically different. As discussed in Refs.~\cite{Skokov:2012ds,BRAUNMUNZINGER2017114}, the initial volume fluctuation can be partly suppressed within the framework of the Wounded Nucleons Model (WNM)~\cite{Bialas:1976ed}. The WNM model assumes that produced particles in nucleus-nucleus collisions are generated from inelastic scattered wounded nucleons. Each wounded nucleon (or participating nucleon) is treated as an independent source and contributes to the total number of produced particles. However, the difference in the $N_{\rm part}$ distributions of UrQMD and Glauber Model would imply a strong model dependence.  

In order to demonstrate the effect of volume fluctuations, a volume fluctuation correction (VFC) method proposed in Ref.~\cite{BRAUNMUNZINGER2017114} has been applied to proton cumulants from the transport UrQMD model. As seen in Fig.~\ref{fig:gm_urqmd_npart}, the $N_{\rm part}$ distributions are different using different centrality determination methods. As a result, there are sizable differences in the corrected proton cumulants results shown in Fig.~\ref{fig:urqmd_vfc_ratio} where black dots are the ratios of proton cumulants from the UrQMD model. The results of the corrected ratios, using $N_{\rm part}$ determined from the UrQMD model directly or from Glauber fits to the charged multiplicity distributions, done exactly as in data analysis in the experiment, are shown by blue squares and red triangles, respectively. Although the correction with the UrQMD $N_{\rm part}$ is supposed to be the answer, the results with Glauber are quite different except for the most central collisions. The maximum effect, due to the volume fluctuations, is around the mid-central centrality bins.  This is qualitatively consistent with the mid-central peak of the dispersion in $N_{\rm part}$ presented in Fig.~\ref{fig:urqmd_npart_width} (b). The negligible impact on the most central Au+Au events is due to the constraint of the total number of participating nucleons $N_{\rm part}^{\rm max}=394$. 

\begin{table}[!htbp]
\center
\begin{tabular}{c|c}
\toprule
Centrality (\%) & $N_{\rm part}\geq$  \\
\midrule
0--5    & 342 \\
5--10   & 307 \\
10--20  & 240\\
20--30   & 180 \\
30--40    & 129\\ 
40--50   &88 \\
50--60  & 55\\
60-70 & 31\\
70-80&15\\
\bottomrule
\end{tabular}
\caption{The centrality definition determined by $N_{\rm part}$ in \auau{} collisions at $\sqrtsNN$ = 3 GeV from UrQMD model. The centrality definition is only used in UrQMD calculation.}
\label{tab:npart_cent}
\end{table}

Calculations from the UrQMD model with the fixed impact parameter range $b\le3$ fm, corresponding to the top 5\% collisions, are performed for the 3 GeV Au+Au collisions. The result is shown as blue open crosses in Fig.~\ref{fig:urqmd_vfc_ratio}. As mentioned earlier, although the selection of centrality with impact parameter or $N_{\rm part}$(Tab.~\ref{tab:npart_cent}) eliminated the IVF in the model calculations, it is not experimentally measurable. In addition, the approach collected a different class of events as shown in Figs.~\ref{fig:urqmd_npart_width} and~\ref{fig:gm_urqmd_npart}.  

\begin{figure*}
	\centering
	\includegraphics[width=0.9\textwidth]{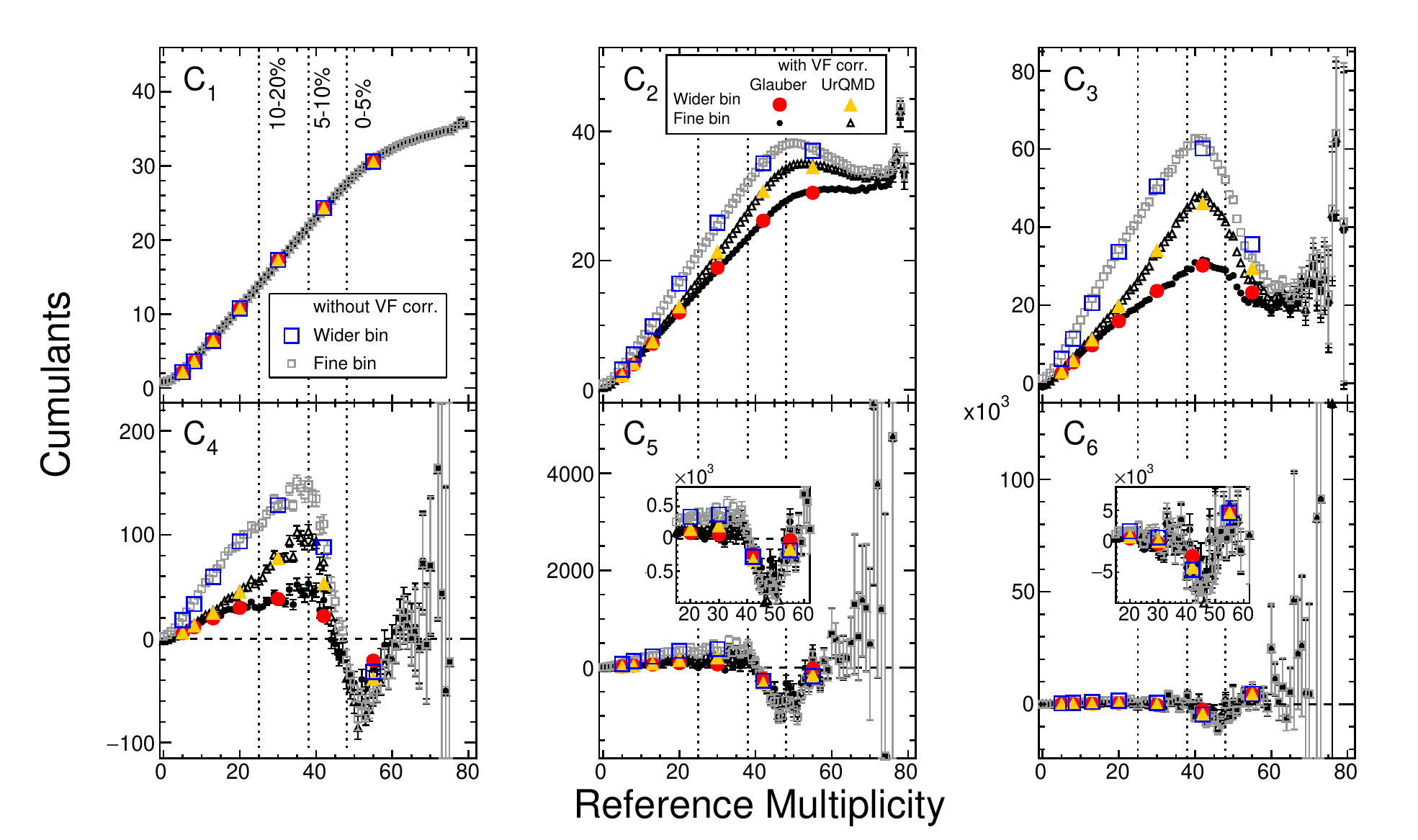}
	\caption{Proton cumulants up to $6^{\rm th}$-order in $\sqrtsNN$ = 3 GeV Au+Au collisions. Data without volume fluctuation correction is shown as grey open squares while data with volume fluctuation correction using $N_{\rm part}$ distributions from Glauber and UrQMD models are shown as black circles and black open triangles, respectively. The corresponding centrality binned cumulants are shown in blue squares, red circles, and orange triangles, respectively. Similar to Fig.~\ref{fig:pu}, the vertical dashed lines indicate the centrality classes.%
}
	\label{fig:vfc_result1} 
\end{figure*}
\begin{figure*}
	\centering
	\includegraphics[width=0.9\textwidth]{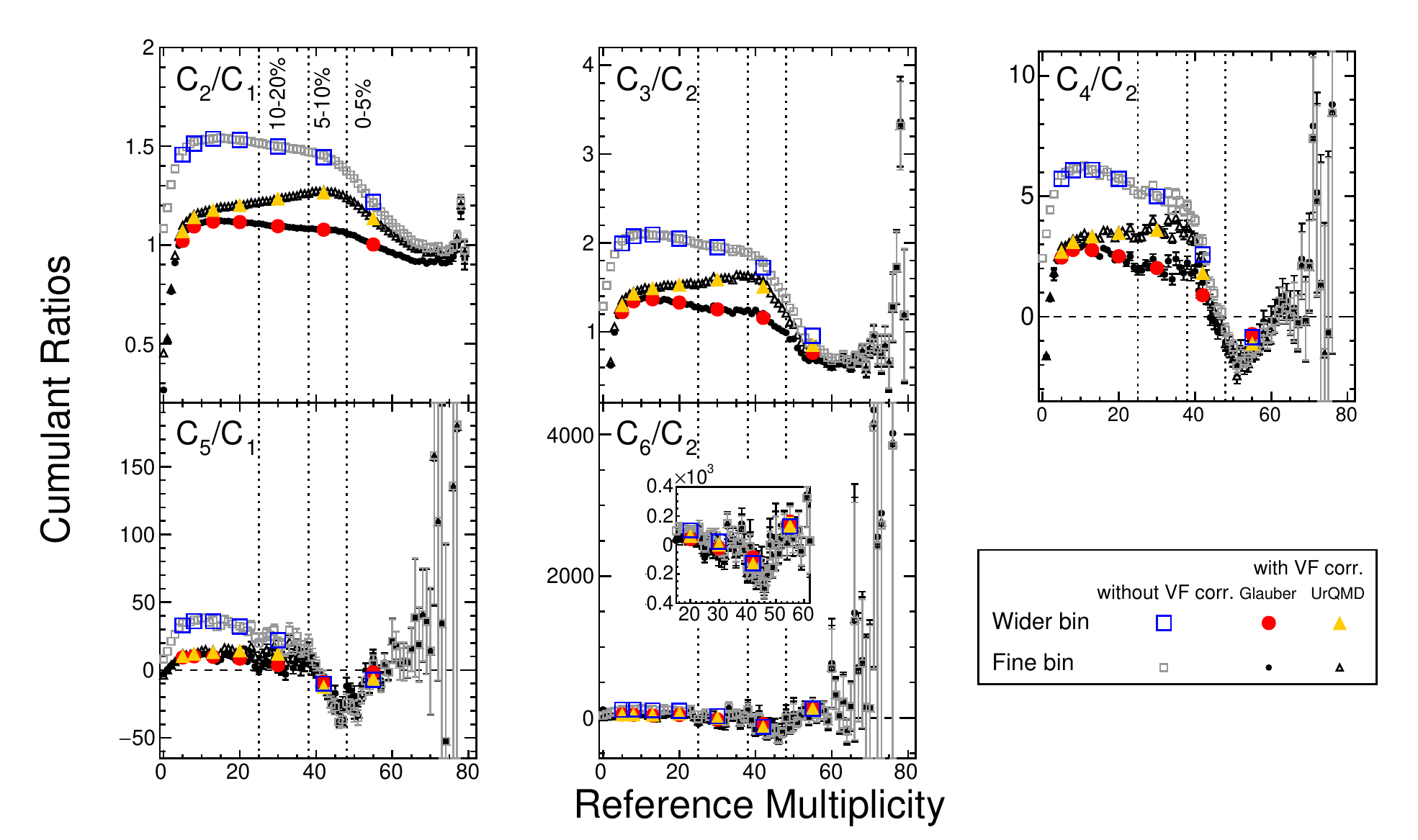}
	\caption{Proton cumulant ratios up to $6^{\rm th}$-order in \mbox{$\sqrtsNN$ = 3 GeV} Au+Au collisions. Data without volume fluctuation correction are shown as grey open squares while data with volume fluctuation correction using $N_{\rm part}$ distributions from Glauber and UrQMD models are shown as black circles and black open triangles, respectively. The corresponding centrality binned cumulants are shown in blue squares, red circles, and orange triangles, respectively. Similar to Fig.~\ref{fig:pu}, the vertical dashed lines indicate the centrality classes.%
}
	\label{fig:vfc_result2} %
\end{figure*}

\begin{figure*}
	\centering
	\includegraphics[width=0.9\textwidth]{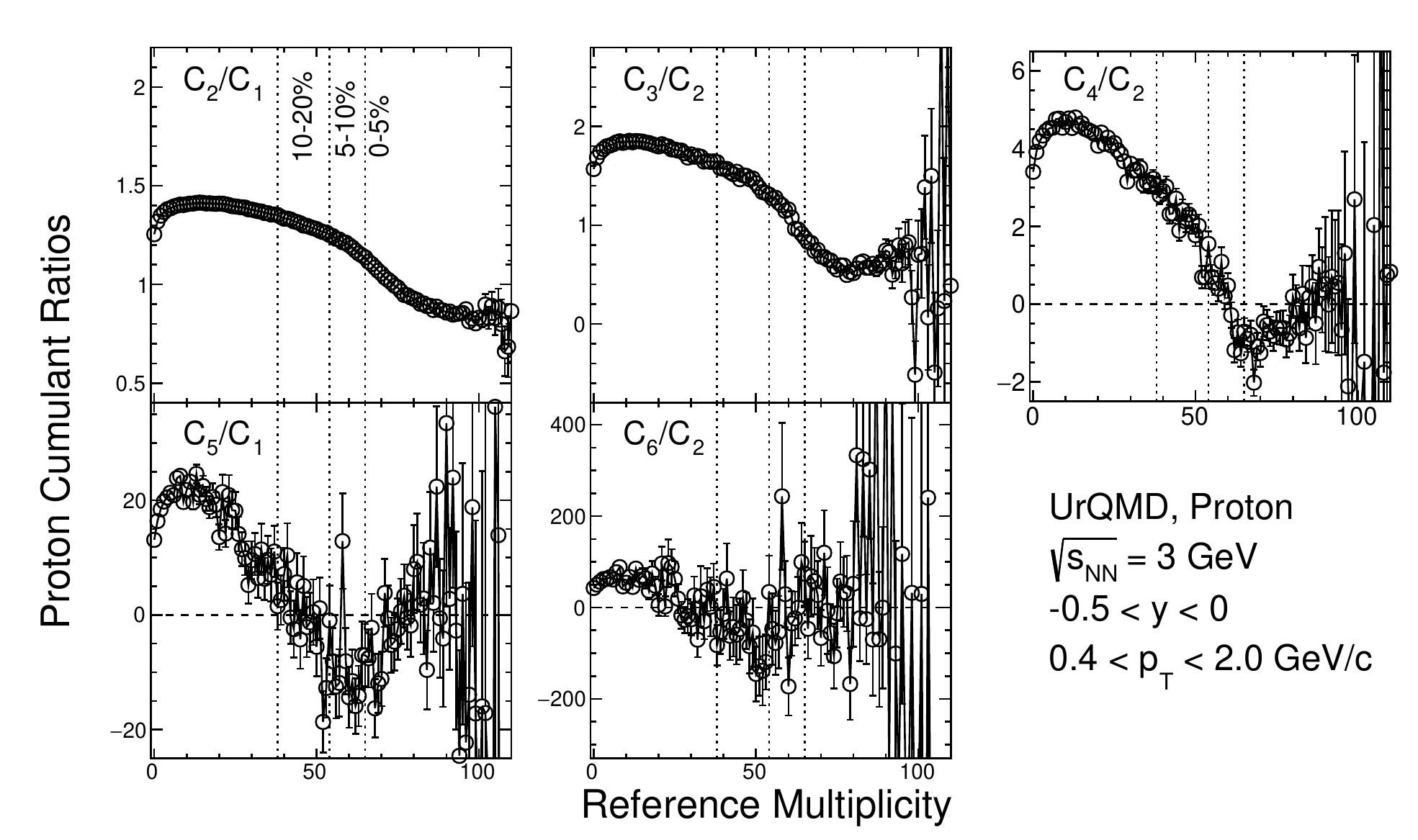}
	\caption{UrQMD results of proton cumulant ratios up to $6^{\rm th}$-order in \auau{} collisions at $\sqrtsNN$ = 3 GeV. The vertical dashed lines indicate the centrality classes.
	}
	\label{fig:urqmd_ref3_dependence2} 
\end{figure*}

%--======================================================================
\subsection{Statistical and systematic uncertainty}

\begin{table}[!htbp]
\center
\begin{tabular}{c|c|c}
\toprule
Source & Nominal &  Variations \\
\toprule
{Centrality}  &$N_{\rm ch}$ & $\pm 1$\\
{Pileup fraction}  & 0.46\%   & 0.37\%, 0.55\% \\
\midrule
TPC spatial hits &   10 & 12 , 15 \\
DCA (cm)  & 3.0 & 2.75, 2.5, 2.0\\
TOF $m^2$ (GeV$^2$/$c^4$)  & (0.6, 1.2) & (0.5, 1.3), (0.7, 1.1)  \\
Efficiency ($\epsilon$)         & $\epsilon$ & $\epsilon\times 1.05$, $\epsilon\times 0.95$  \\
\bottomrule
\end{tabular}
\caption{Sources, choices of nominal values, and their variations for systematic uncertainties in proton cumulant measurements from the fixed-target Au+Au collisions at \mbox{$\sqrtsNN$ = 3\,GeV}. The nominal values of $N_{\rm ch}$ can be seen Table~\ref{tab:sys_plot}. }
\label{tab:systematic_plot}
\end{table}

\begin{table*}%[!htbp]
	\setlength{\tabcolsep}{0.30cm}{
	\center
	\begin{tabular}{c|cccccc}
	\toprule
	Source     & $C_2/C_1$  & $C_3/C_2$ &$C_4/C_2$ &$C_5/C1$&$C_6/C_2$   \\
	& 1.218$\pm$0.001  & 0.954$\pm$0.005            & -0.845$\pm$0.086  &-7.104$\pm$2.163&128.752$\pm$51.401  \\
	\midrule
	Centrality        & 0.014  & 0.041  & 0.042 &2.330&23.967\\
	Pileup            & 0.002  & 0.017  & 0.242&3.519&50.990\\
	\midrule
	TPC hits          & 0.002  & 0.015 & 0.241&5.334&115.492\\
	DCA               & 0.008  & 0.037 & 0.784 &15.688&302.049\\
	TOF $m^2$         & 0.003  & 0.009 & 0.050 &0.643&10.324\\
	Efficiency  & 0.011 & 0.023 & 0.272&0.277&49.774\\
	\midrule
	Total & 0.018   & 0.058 &  0.822 &16.246&307.259\\
	\bottomrule
	\end{tabular}}
	\caption{Main contributors to systematic uncertainty to the proton cumulant ratios: $C_2/C_1$, $C_3/C_2$, and $C_4/C_2$ from 0-5\% central 3\,GeV Au+Au collisions. The first row shows the values and statistical uncertainty of those ratios. The corresponding values of these ratios along with the statistical uncertainties are listed in the table. The final total value is the quadratic sum of uncertainties from centrality, pileup, and the dominant contribution from TPC hits, DCA, TOF $m^{2}$, and detector efficiency. Clearly, this analysis is systematically dominant.}
	\label{tab:uncertainty}
\end{table*}

The statistical uncertainties are obtained using the Bootstrap approach~\cite{bootstrap_method} in which events are re-sampled with replacement and the analysis is re-run. The Bootstrap procedure is repeated 200 times and the statistical uncertainty is the standard deviation of the Bootstrapped observable values, such as the cumulants and their ratios. 

The systematic uncertainty of the cumulant calculation can be subdivided into three categories: pileup correction, centrality determination (Tab.~\ref{tab:sys_plot}), and track selection. The track selection includes the track reconstruction requirements (TPC spatial hits, DCA), the mass-squared cut, and the efficiency in the TPC and TOF. 
 The effect of lowering the ${\rm d}E/{\rm d}x$ cut to $|N_{\sigma,p}|<2$ was tested but did not affect the final result. 

To estimate the systematic uncertainty, the analysis was repeated with different analysis requirements which are outlined in Table~\ref{tab:systematic_plot}.
The final total value is the quadratic sum of uncertainties from centrality, pileup, and the dominant contribution from TPC points, DCA, and PID $m^{2}$ and efficiency $\epsilon$. 
The difference between the systematic analyses and nominal analysis in $C_2/C_1$, $C_3/C_2$, and $C_4/C_2$ in 0-5\% central Au+Au collisions is listed in Table \ref{tab:uncertainty}. 
%---===================================================================
\section{Results and discussions}
\label{sec:result}
\subsection{Experimental Results}
\label{sec:resutls1}
Experimental data of proton cumulants and their ratios as a function of the reference multiplicity from 3 GeV \auau{} collisions are shown in Fig.~\ref{fig:vfc_result1} and Fig.~\ref{fig:vfc_result2}. 
%Results with and without the volume fluctuation corrections are shown as black circles and open blue squares, respectively. 
The reference multiplicity dependence of data without initial volume fluctuation correction is shown as grey open squares while data with volume fluctuation correction using $N_{\rm part}$ distributions from Glauber and UrQMD model are shown as black filled circles and black open triangles, respectively. The centrality binned results with CBWC are shown as blue open squares, red filled circles and orange filled triangles correspondingly. By definition, $C_1$ is not affected by the volume fluctuation correction while strong model dependence for higher order cumulants in the initial volume fluctuation corrections is clear in the figure.
 
For higher-order cumulants, a maximum difference between results with and without VFC is seen around mid-central \auau{} collisions and the difference slightly depends on the order of the cumulants. In the most central bin, the corrected and uncorrected proton cumulants $C_{i}$ ($i>3$) are very similar. One can see that cumulants show strong multiplicity dependence. Rapid decreases are seen from mid-central (5-10\%) to most central collisions (0-5\%) in $C_3$ to $C_6$. And in the high reference multiplicity region ($>50$) there is an increase with multiplicity. Indeed, as one can see that in the central collision region, 0-5\% and 5-10\%, the values of $C_4$ to $C_6$ (Fig.~\ref{fig:vfc_result1}) and corresponding ratios (Fig.~\ref{fig:vfc_result2}) change from positive to negative and to positive value again at the multiplicity larger than 60. Again, when multiplicity is larger than 50, the VFC shows no effect on either the proton cumulants ($C_4, C_5$ and $C_6$ in Fig.~\ref{fig:vfc_result1}) or their ratios ($C_4/C_2, C_5/C_1$, and $C_6/C_2$ in Fig.~\ref{fig:vfc_result2}).

In later discussions, the experimental data will be compared with model calculations. For the UrQMD model, around 80 million events are produced in \auau{} collisions at $\sqrtsNN$ = 3 GeV with the cascade mode. The same kinematic cuts and centrality bins used in the data analysis are applied in the model calculations. The collision centrality is determined using the charged particle multiplicity excluding protons in the acceptance of the TPC, the same procedure as used in the data analysis. The CBWC procedure is also applied to get the properly weighted centrality binned model results.

The UrQMD model calculations as a function of reference multiplicity are shown in Fig.~\ref{fig:urqmd_ref3_dependence2}. Cumulants show a strong dependence on reference multiplicity.
The strong multiplicity dependence in the higher order of the proton ratios is very similar to that observed in experimental data, see Fig.~\ref{fig:vfc_result2}. Stronger variations are seen in the higher order ratios.

\begin{figure*}
	\centering
	\includegraphics[width=0.8\textwidth]{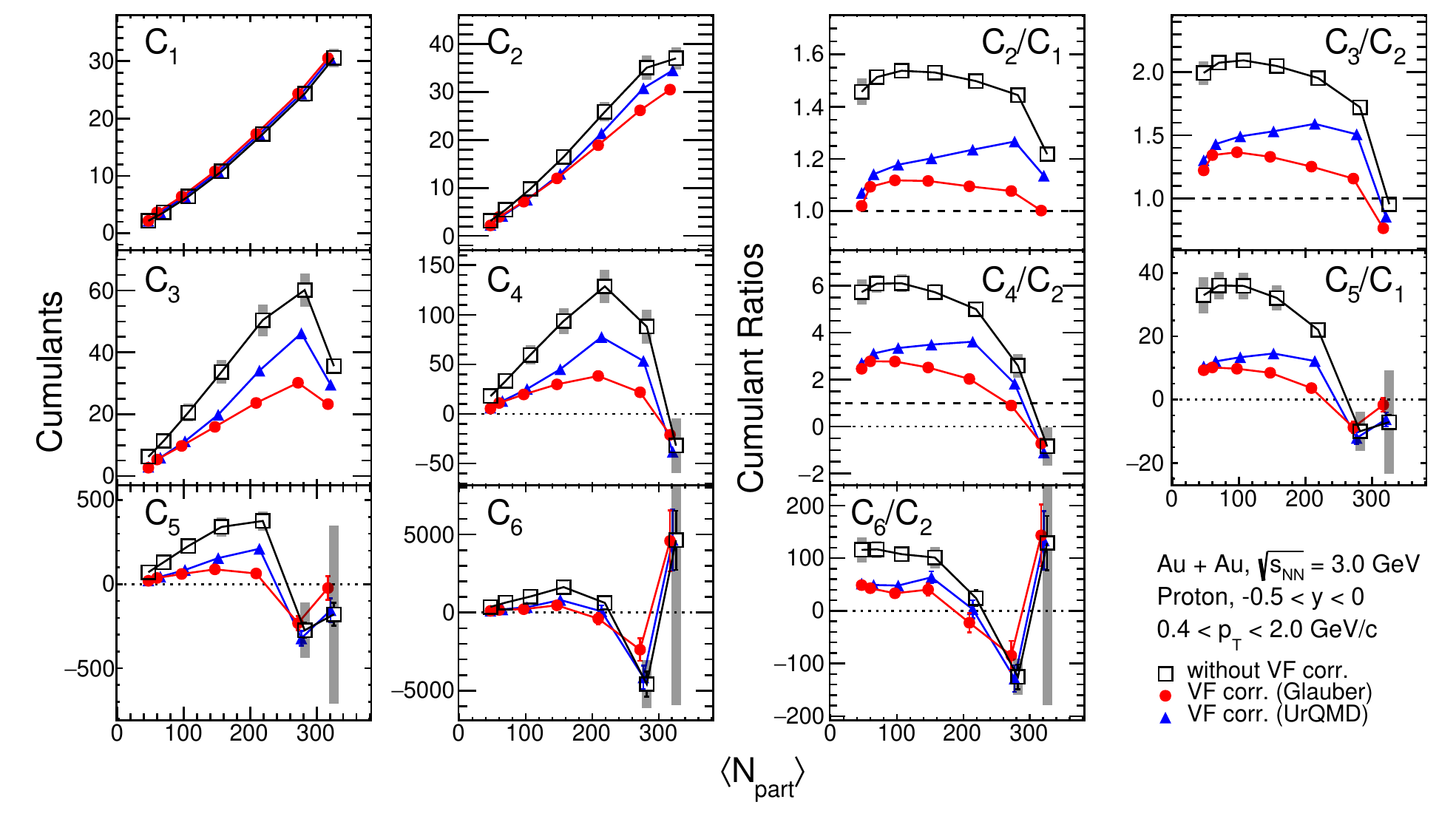}
	\caption{Experimental results on centrality dependence of cumulants (left panels) and cumulant ratios (right panels) up to $6^{\rm th}$-order of the proton multiplicity distributions in \auau{} collisions at $\sqrtsNN$ = 3 GeV. The open squares are data without VF correction while red circles and blue triangles are results with VF correction with $N_{\rm part}$ distributions from Glauber and UrQMD models, respectively. }
	\label{fig:cumulant_vfc}
\end{figure*}

\begin{figure*}
	\centering
	\includegraphics[width=0.8\textwidth]{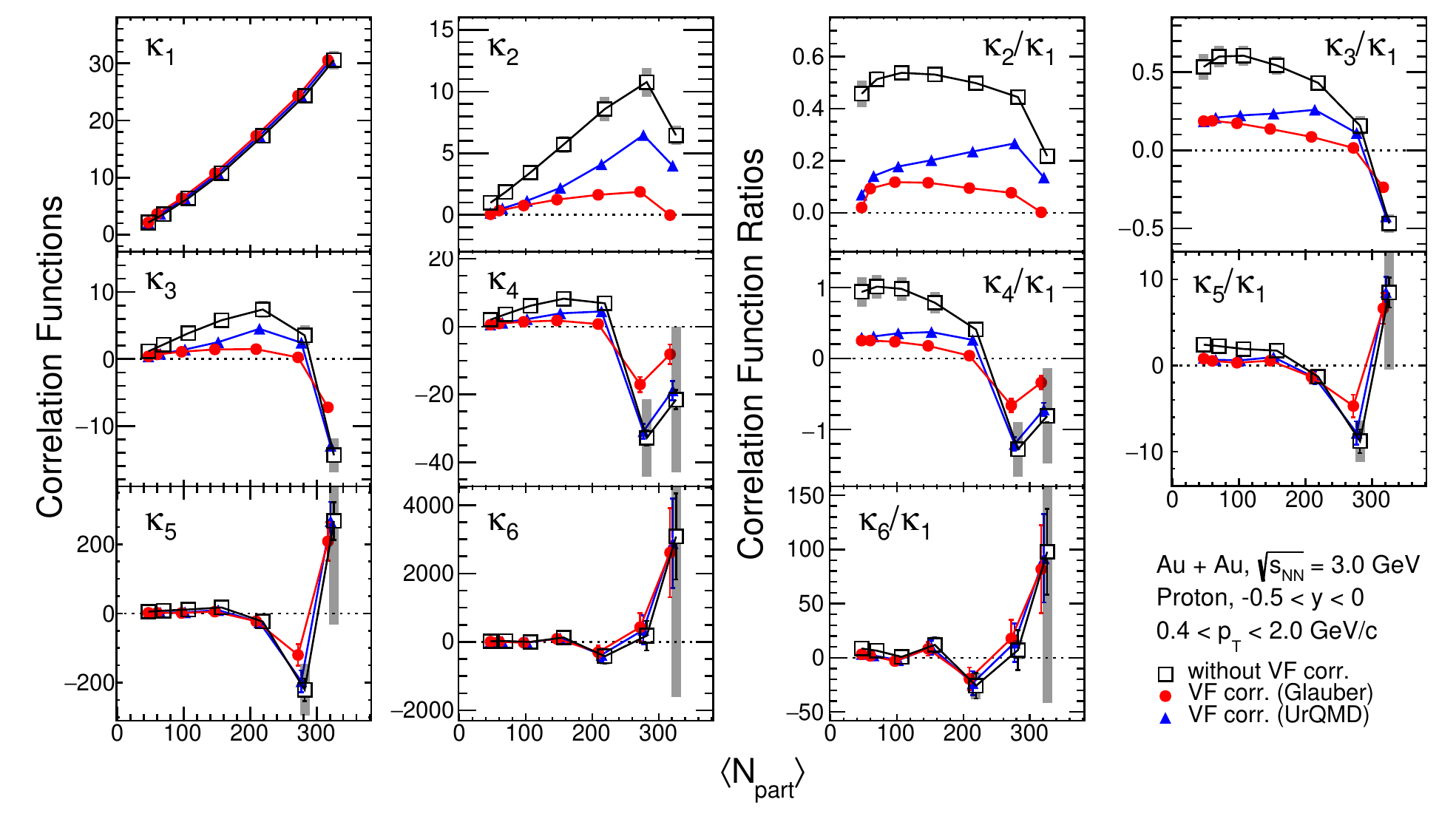}
	\caption{Same as Fig.~\ref{fig:cumulant_vfc} but for correlation function (left panels) and their normalized ratios (right panels).}
	\label{fig:correlation_function_vfc}
\end{figure*}

%--===========================================================
\subsection{Collision centrality dependence}
In this section, we discuss the centrality dependence of the proton cumulants and correlation functions along with the corresponding ratios. Assuming that collisions are a superposition of independent sources, one expects the cumulant values to increase with $\mean{N_{\rm part}}$. The centrality classes are related to the average number of participating nucleons $\mean{N_{\rm part}}$ shown in Tab.~\ref{tab:sys_plot}.

\begin{figure*}
	\centering
	\includegraphics[width=0.8\textwidth]{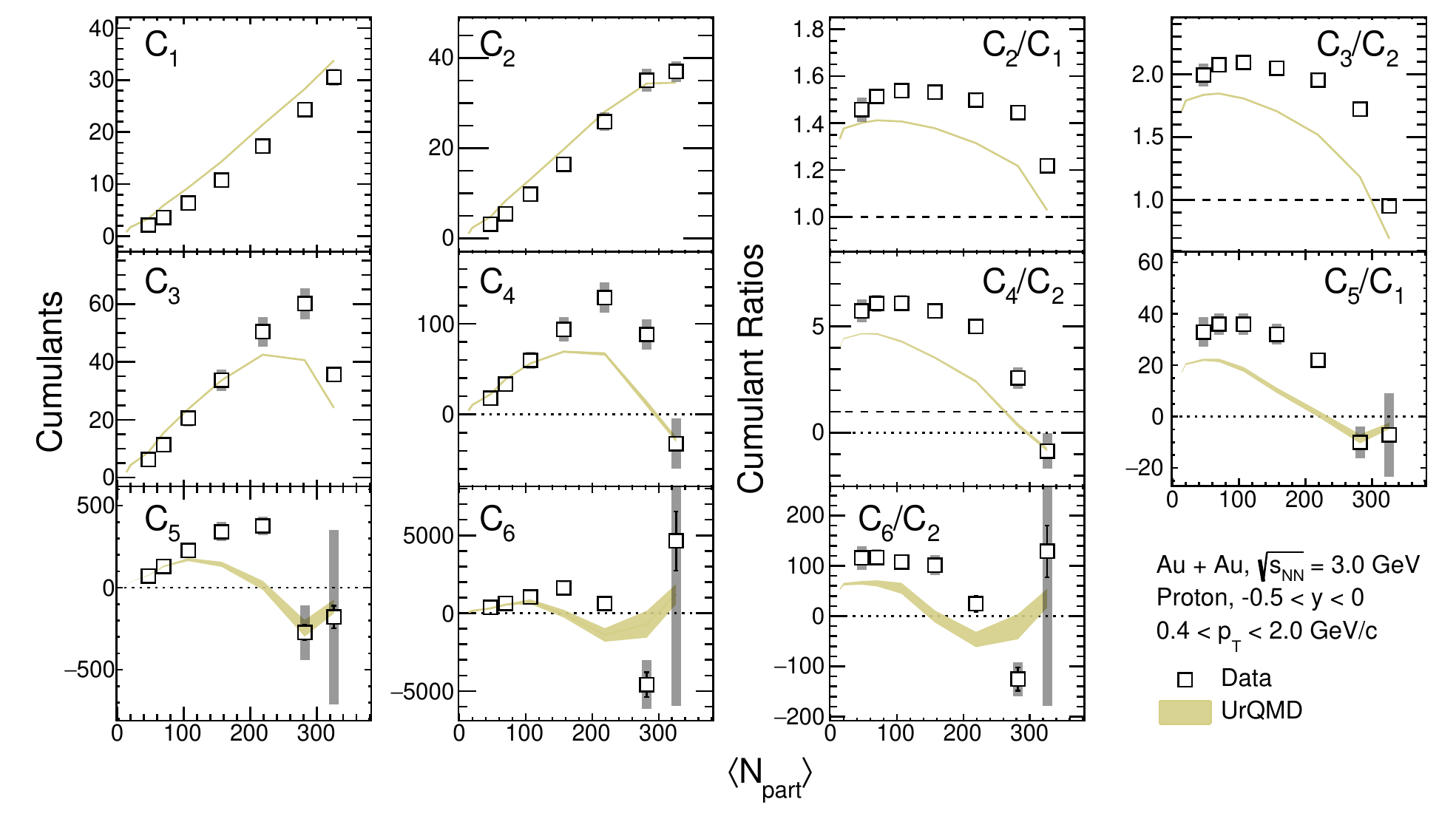}
	\caption{Cumulants and cumulant ratios of proton multiplicity distributions for \auau{} collisions at $\sqrtsNN$ = 3 GeV. The transverse momentum window is $\ppt$ from $0.4 < p_{\rm T} < 2.0$ GeV/$c$ and the rapidity window is $-0.5 < y < 0$. Statistical and systematic uncertainties are represented by black and gray bars, respectively. UrQMD predictions are depicted by gold bands.}
	\label{fig:cumu_cent}
\end{figure*}

Figures~\ref{fig:cumulant_vfc} and \ref{fig:correlation_function_vfc} show proton cumulants and correlation functions as a function of $\mean{N_{\rm part}}$ with VFC using two different models. In the cumulant and correlation function ratios, the corrections suppress the large values of cumulant ratios in mid-central and peripheral collisions. In addition, one observes large variations in the VF corrected results between the UrQMD and Glauber Model. However, as can be seen in these figures, in the most central 0-5\% \auau{} collisions, the difference for higher order ones ($C_{i},\quad i>2$) between data without the VFC and with the VFC using different model inputs is small. This implies that the results of cumulants as well as the correlation functions in the most central collisions are least affected by the volume fluctuations. Because of the strong model dependence, starting from Fig.~\ref{fig:cumu_cent} the VFC method is not adopted and the results from 3 GeV experimental data are only applied with pileup correction and CBWC. For UrQMD results, only CBWC is applied.  

\begin{figure*}
	\centering
	\includegraphics[width=0.8\textwidth]{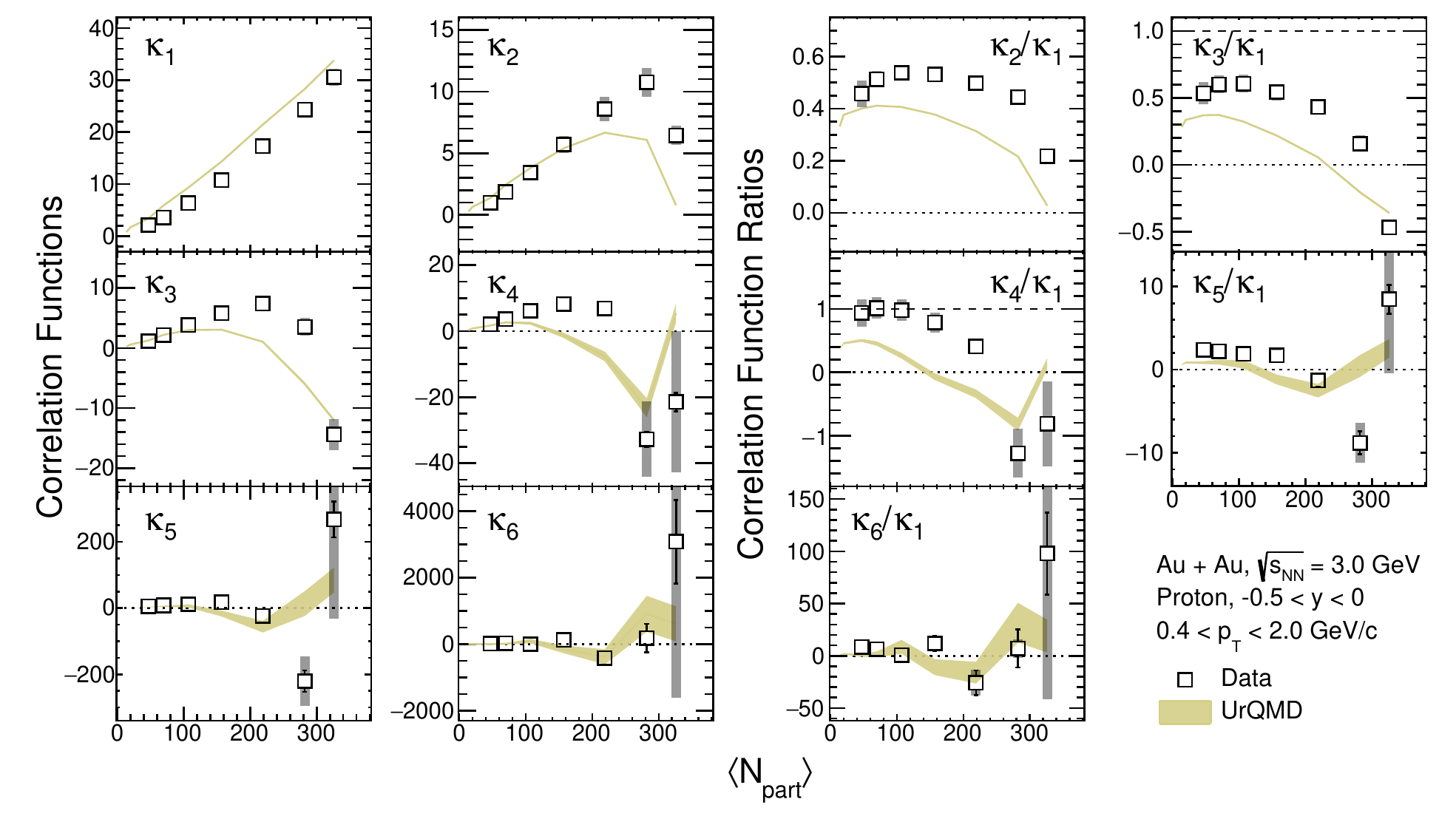}
	\caption{Same as Fig.~\ref{fig:cumu_cent} but for correlation functions and correlation function ratios of proton multiplicity distributions for Au+Au collisions at $\sqrtsNN$ = 3 GeV.}
	\label{fig:corr_cent}
\end{figure*}

Figure~\ref{fig:cumu_cent} shows the centrality dependence of the proton cumulants and their ratios extracted from the kinematic acceptance $-0.5<y<0$ and $0.4<\ppt<2.0$ GeV/$c$. The experimental data are compared with the UrQMD calculations (gold bands). As one can see, only $C_1$ and $C_2$ values increase with $\mean{N_{\rm part}}$. For the higher order cumulants ($C_{i},\quad i>2$), the cumulants increase with $\mean{N_{\rm part}}$ ($ \mean{N_{\rm part}} < 200$) but change rapidly in the more central centrality region. For all cumulant ratios, the values are above unity in the peripheral and are closer to unity for mid-central collisions. Figure~\ref{fig:corr_cent} shows the centrality dependence of the proton correlation functions and ratios with the same acceptance as Fig.~\ref{fig:cumu_cent}. The correlation functions also deviate from a monotonic increase around $\mean{N_{\rm part}} \sim 200$. 
The trends of cumulants and correlation functions shown in Figs.~\ref{fig:cumu_cent} and \ref{fig:corr_cent} can be qualitatively reproduced by the UrQMD calculations.

%--======================================================================
\subsection{Rapidity and $p_{\rm T}$ dependence}

In this section, we discuss the rapidity and $p_{\rm T}$ dependence of the cumulants and cumulant ratios.

\begin{figure*}[htbp]
	\centering
	\includegraphics[width=0.8\textwidth]{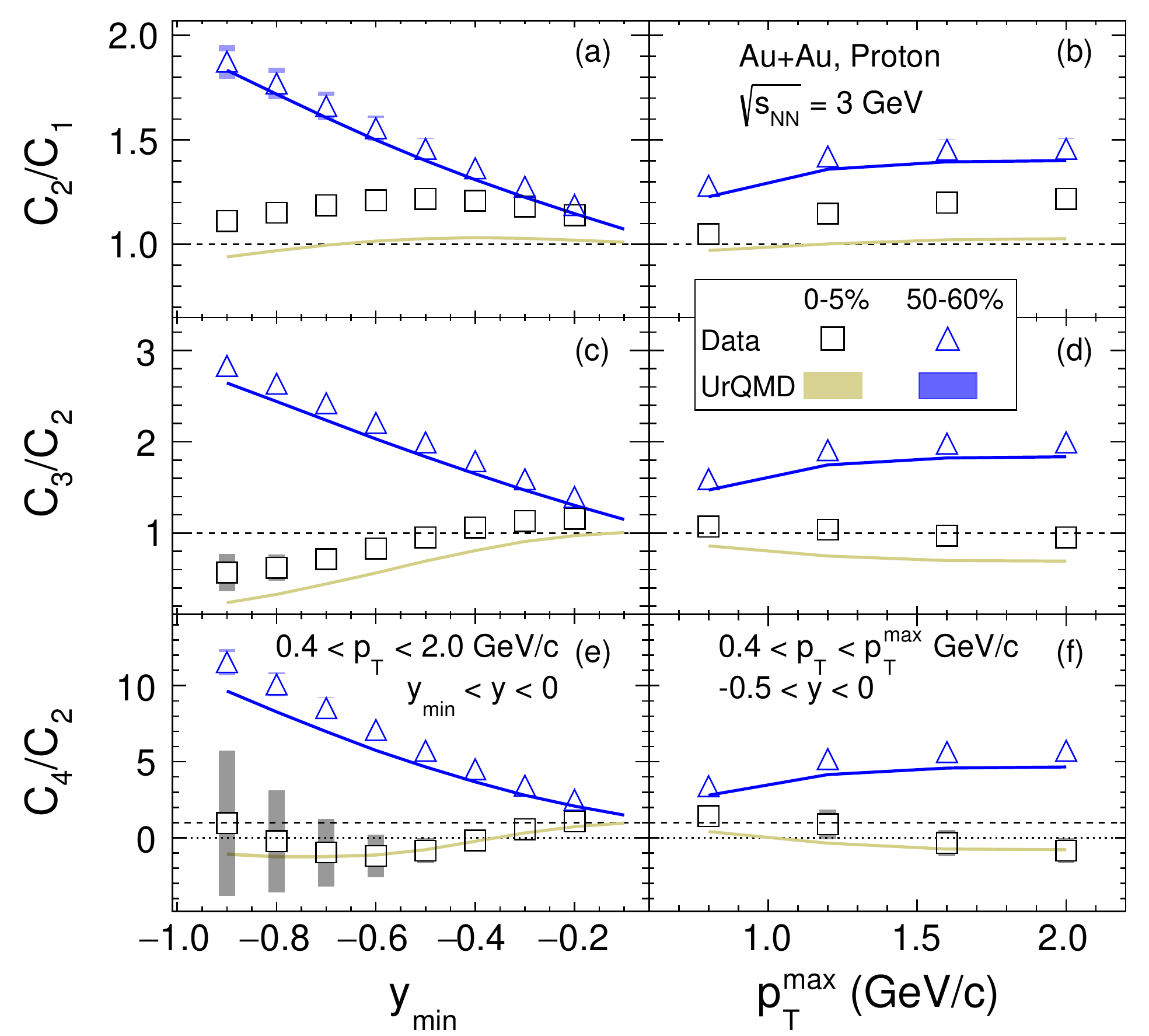}
	\caption{The transverse-momentum and rapidity dependence of cumulant ratios of proton multiplicity distributions for \auau{} collisions at $\sqrtsNN$ = 3 GeV. In the left column, the $\ppt$ of the analysis window is $0.4 < p_{\rm T} < 2.0$ GeV/$c$ while the rapidity window is varied in the range $y_{\rm min} < y < 0$. In the right column, the rapidity of the analysis window is $-0.5 < y < 0$ while the $\ppt$ is varied in the range $0.4 < p_{\rm T} < p_{\rm T}^{\rm max}$ GeV/$c$. The most central (0-5\%) and peripheral (50-60\%) events are depicted by black squares and blue triangles, respectively. Statistical and systematic uncertainties are represented by black and gray bars, respectively. UrQMD simulations for the top 0-5\% and 50-60\% are shown by gold and blue bands, respectively.}
	\label{fig:cumu_ratios}
\end{figure*}

\begin{figure*}[htbp]
	\centering
	\includegraphics[width=0.8\textwidth]{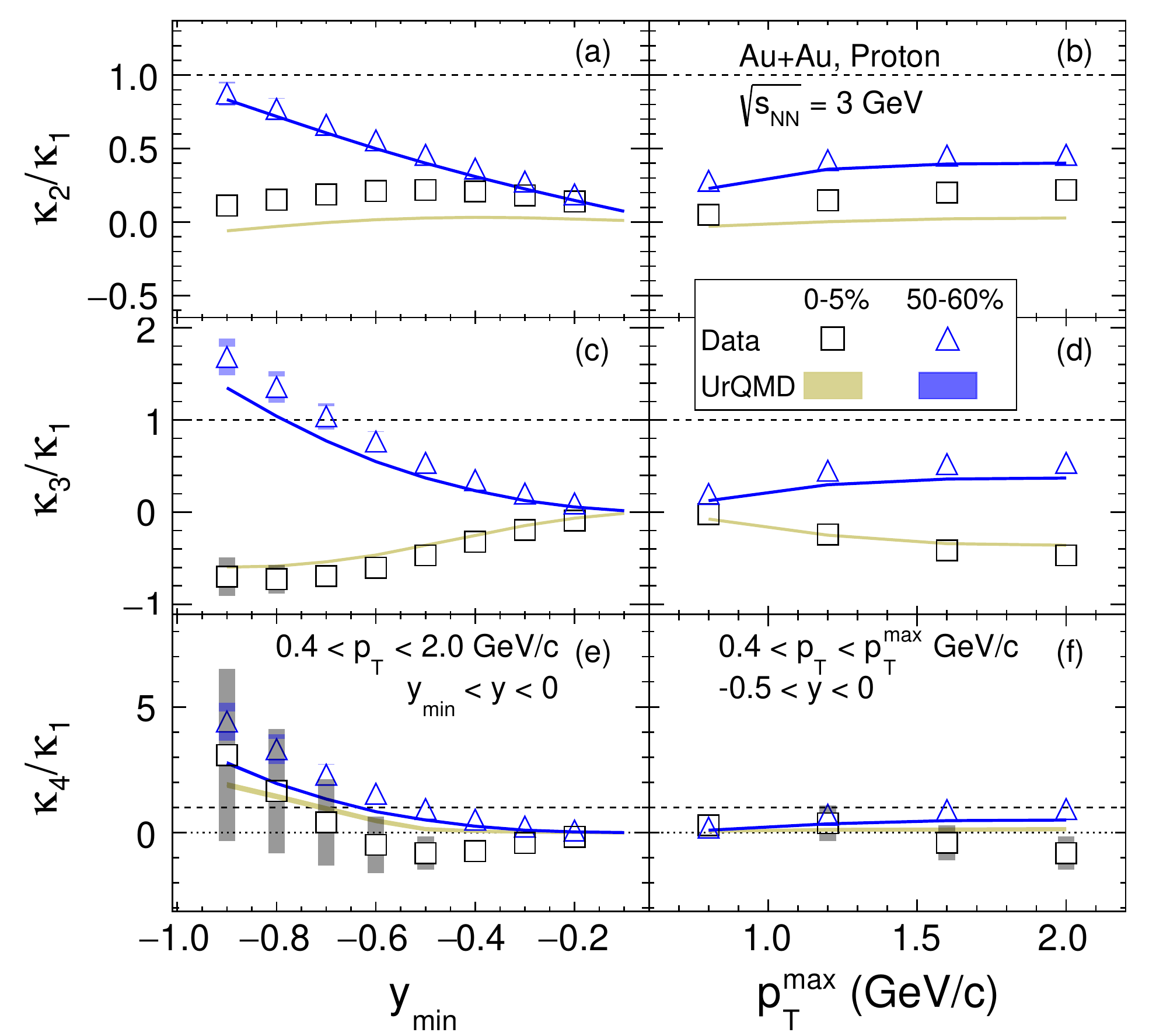}
	\caption{As Fig.~\ref{fig:cumu_ratios} but for transverse-momentum and rapidity dependence of correlation function ratios of proton multiplicity distributions for \auau{} collisions at $\sqrtsNN$ = 3 GeV.}
	\label{fig:corr_ratios}
\end{figure*}

Figures~\ref{fig:cumu_ratios} and \ref{fig:corr_ratios} depict the rapidity (left panels) and transverse momentum (right panels) dependence of ratios of proton cumulants and correlation functions. Data from most central ($0-5\%$) and peripheral ($50-60\%$) centrality classes are shown in the figures. The measured rapidity window covers $y_{\rm min}<y<0$, where $y_{\rm min}$ changes from -0.2 to -0.9 and $\ppt$ window is $0.4<\ppt$ (GeV/$c)<\ppt^{\rm max}$, where $\ppt^{\rm max}$ is varied from $0.8$ to $2.0$ GeV/$c$. Corresponding results from the UrQMD calculations are shown as colored bands in the figures.

As one can see, in the most central collisions, the cumulant ratio $C_2/C_1$ in Fig.~\ref{fig:cumu_ratios}, remains above unity at all rapidities. 
The $C_3/C_2$ ratio is slightly above unity for the smallest rapidity window ($y_{\rm min}=-0.1$) and decreases as the rapidity window increases.

As is expected, the
$C_4/C_2$ ratio is close to unity in the smallest rapidity window and
seems to go back to unity with large uncertainty when the rapidity window is larger than $y$ = -0.5. Similarly, the ratios of the correlation function in Fig.~\ref{fig:corr_ratios} (e) are also close to zero (Poisson baseline) at the smallest rapidity window but show deviations from zero when it goes to a larger rapidity window. These rapidity dependencies are reproduced by the UrQMD calculations.

Overall, the results of the hadronic transport model UrQMD calculations qualitatively reproduce both the rapidity and transverse-momentum dependence. As discussed earlier, in addition to the genuine collision dynamics in the model, the effect of the volume fluctuations is also present in the calculation.

\subsection{Collision energy dependence of cumulant ratios}
%--===========================================================

\begin{figure*}[htbp]
	\centering
	\includegraphics[width=0.6\textwidth]{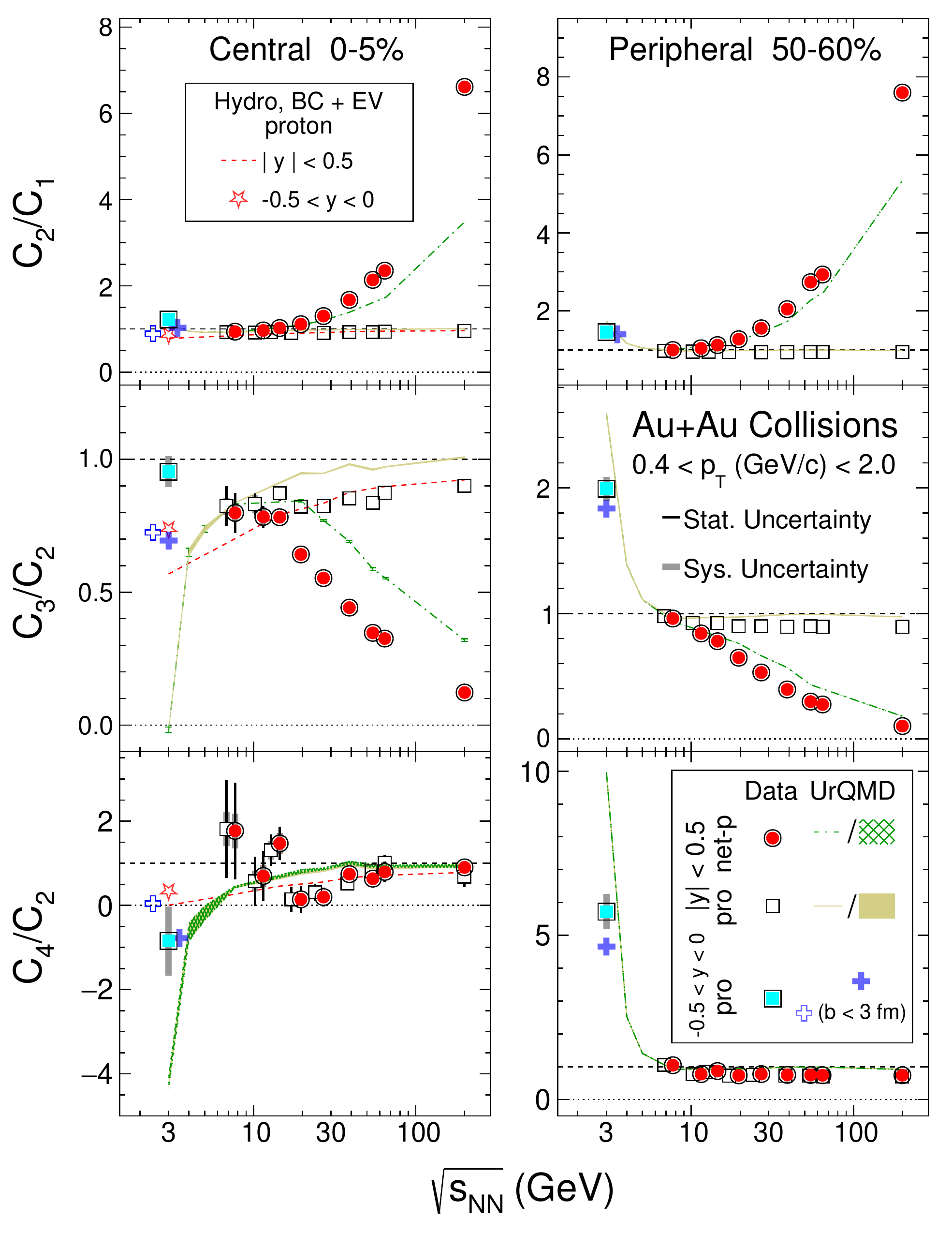}
	\caption{Collision energy dependence of the cumulant ratios: $C_2/C_1 = \sigma/M$, $C_3/C_2 = S\sigma$, and $C_4/C_2= \kappa\sigma^2$, for proton (open squares) and net-proton (red-circles) from top 0-5\% (top panels) and 50-60\% (bottom panels) \auau{} collisions at RHIC. The points for protons are shifted horizontally for clarity. The new result of proton from $\sqrtsNN$ = 3 GeV \auau{} collisions is shown as filled square. UrQMD results with $|y| < 0.5$ of proton are shown as gold bands while that of net-proton are shown as green dashed lines or green bands. At 3 GeV, the model result for proton ($-0.5 < y < 0$) are shown as blue crosses. UrQMD results of proton and net-proton $C_4/C_2$, see right panels, are almost totally overlapped. The open cross is the result of the model with a fixed impact parameter $b<3$ fm. The hydrodynamic calculations, for 5\% central \auau{} collisions, for protons from $|y|<0.5$ are shown as dashed red line and the result of the 3 GeV protons from $-0.5<y<0$ is shown as open red star.  }
	\label{fig:energy_dependence_cumulant}
\end{figure*}

In the first order, taking the ratio of cumulants cancels the effect of volume but not the fluctuations in volume. Figure~\ref{fig:energy_dependence_cumulant} depicts the collision energy dependence of the cumulant ratios from 0-5\% central (top panels) and 50-60\% peripheral (bottom panels) collisions. The new result of protons from 3 GeV Au+Au collisions data, shown as filled squares, is compared to that of protons (open squares) and net-protons (filled circles) from \auau{}  collisions at \mbox{$\sqrtsNN$ = 7.7 -- 200 GeV}.

UrQMD results with $|y| < 0.5$ of proton are shown as gold bands while that of net-proton are shown as green dashed lines or green bands. At 3 GeV, the model result for proton ($-0.5 < y < 0$) are shown as blue crosses.      
While the net-proton ratios show a clear energy dependence, the proton $C_2/C_1$ and $C_3/C_2$ ratios are relatively flat and around unity as a function of collision energy except for the 3 GeV data.
The new proton data from 3 GeV does not follow this trend in the most central collisions.

Notably, both proton (open squares) and net-proton (filled circles) cumulant ratios converge at collision energies below 20 GeV. This implies that at high baryon density region, the anti-proton production becomes negligible. At the center of mass energy of 2.4 GeV, HADES reported the values for proton cumulant ratios in 0-10\% central \auau{} collisions: \mbox{$C_3/C_2$ = -1.63 $\pm$ 0.09 (stat) $\pm$ 0.34 (sys)} and \mbox{$C_4/C_2$ = 0.15 $\pm$ 0.9 (stat) $\pm$ 1.4 (sys)} from \mbox{$|y|< 0.4$}, \mbox{$0.4 < p_{\rm T} <1.6$ GeV/$c$~\cite{HADES_Proton}}. While the value of $C_4/C_2$ from the HADES experiment is consistent with the new 3 GeV data, the sign of $C_3/C_2$ is opposite to what we observed here.

Except for the $C_3/C_2$ ratio from 3 GeV central collisions, the UrQMD results reproduce the energy dependence trend well for both proton and net-proton, see green and gold bands in the figure. 
For the peripheral 50-60\% collisions, the $C_4/C_2$ ratio from 3 GeV is larger than that from higher energy collisions, by a factor of five. A rapid increase in the energy dependence seems confirmed by the UrQMD model calculations, see both the blue cross and gold band in the figure. In the 3 GeV most central collisions, unlike all higher energy collisions, the value of $C_4/C_2$ is negative. The UrQMD model calculations, again, reproduce the trend well: due to baryon number conservation, the $C_4/C_2$ is dramatically suppressed in the high baryon density region.  

Hydrodynamic calculations are shown as red dashed lines in Fig.~\ref{fig:energy_dependence_cumulant} for the 0-5\% \auau{} collisions.
The hydrodynamic evolution is made with the open-source code MUSIC (v3.0)~\cite{Denicol:2018wdp}. The initial condition is taken from Ref.~\cite{Shen:2020jwv} and the particlization is given by the Cooper-Frye formula~\cite{Cooper:1974mv} with non-ideal hadron resonance gas model~\cite{Vovchenko:2017xad}. At the grand canonical limit, including both effects of excluded volume and global baryon number conservation, the net-proton cumulants are evaluated on the Cooper-Frye hypersurface. One may find more details of the model calculations in Ref.~\cite{Vovchenko:2021kxx}. Unlike the commonly used transport model approach, here all calculations, starting from initial condition to hydro-evolution to hadronization, are all performed using averaged ensembles.
Cumulant ratios $C_2/C_1$, $C_3/C_2$, and $C_4/C_2$ in hydrodynamic calculations are all below unity. Interestingly, the UrQMD results with a fixed impact parameter are also suppressed, see open blue crosses. Qualitatively, the results from the fixed impact parameter ($b\le 3$ fm blue open crosses) UrQMD calculations follow that of the hydrodynamic calculations (red dashed lines) with a canonical ensemble.

The fact that the negative $C_4/C_2$ in the most central Au+Au collisions at 3 GeV is reproduced by the hadronic transport UrQMD model although $C_3/C_2$ is over-predicted, implies that the system is dominated by hadronic interactions. This conclusion is also consistent with the measurements of the collectivity of light hadrons~\cite{STAR:2021yiu} as well as the strange hadron production~\cite{STAR:2021hyx} at the same collision energy.  

Due to baryon number conservation, proton multiplicity distributions are also modified by the formation of light nuclei in the same collision~\cite{Feckova:2015qza}. The effect is especially strong in the high baryon density region where the production of light nuclei is expected to be relatively high~\cite{Andronic:2010qu}. The influence of the light nuclei on the proton cumulants and their ratios will be analyzed in the future when the yields of light nuclei become available.   

\section{Summary and outlook}
\label{sec:summary}

In summary, we report a systematic measurement of cumulants and correlation functions of proton multiplicities up to the $6^{\rm th}$-order in \auau{} collisions at $\sqrtsNN$ = 3 GeV. The data were collected with the STAR fixed-target mode in the year 2018 at RHIC. The analysis includes the centrality, rapidity, $p_{\rm T}$, and energy dependence of these fluctuation observables for proton multiplicities. Other important effects which are relevant to low-energy fixed-target collisions such as pileup and volume fluctuations are also discussed.  

The protons are identified using the STAR TPC and TOF with purity greater than 95\%. The centrality selection is based on pion and kaon multiplicities in the full acceptance of the TPC. The proton tracks are corrected for detector efficiencies using a binomial response function. The cumulant values are corrected for pileup contamination. The pileup fraction is determined to be (0.46 $\pm$ 0.09)\% for all events and (2.10 $\pm$ 0.40)\% in the 0–5\% centrality class. 

Due to a weak correlation between the measured reference multiplicity and the initial number of participants, a considerable effect from the initial volume fluctuations is expected. Except for the most central collisions, the effects can be suppressed by implementing a model-dependent correction procedure ~\cite{BRAUNMUNZINGER2017114}, however, the results are dependent on the choice of model that provides $N_{\rm part}$ input for the correction procedure. Interestingly, higher-order cumulant ratios $C_4/C_2$, $C_5/C_1$, and $C_6/C_2$ in most central events appear least affected by volume fluctuations in the 3 GeV Au+Au collisions. 

The rapidity, transverse momentum, and centrality dependence are shown for the proton cumulants and their ratios. The UrQMD model reproduces the trends well, however, it does not agree within uncertainties. Compared with data from higher energy collisions, the \mbox{$\sqrtsNN$ = 3 GeV} 
cumulant ratios $C_2/C_1$, $C_3/C_2$, and $C_4/C_2$, except $C_3/C_2$ in central collisions, are well reproduced by UrQMD calculations. This is attributed to effects from volume fluctuations and hadronic interactions. On the other hand, the data and results of both UrQMD and hydrodynamic models of $C_4/C_2$ in the most central collisions are consistent which signals the effects of baryon number conservation and an energy regime dominated by hadronic interactions. Therefore, the QCD critical point, if discovered in heavy-ion collisions, could only exist at energies higher than 3 GeV.

New data sets have been collected during the second phase of the RHIC beam energy scan program for Au+Au collisions at \mbox{$\sqrtsNN$ = 3 -- 19.6 GeV}. The data sets will have extended kinematic coverage and higher statistics. This will allow to reduce the statistical uncertainties significantly and expand the systematic analysis of both $p_{\rm T}$ and rapidity dependence to wider regions. These studies will be crucial in exploring the QCD phase structure in the high baryon density region and locating the elusive critical point.    

\begin{acknowledgments}
We thank the RHIC Operations Group and RCF at BNL, the NERSC Center at LBNL, and the Open Science Grid consortium for providing resources and support.  This work was supported in part by the Office of Nuclear Physics within the U.S. DOE Office of Science, the U.S. National Science Foundation, National Natural Science Foundation of China, Chinese Academy of Science, the Ministry of Science and Technology of China and the Chinese Ministry of Education, the Higher Education Sprout Project by Ministry of Education at NCKU, the National Research Foundation of Korea, Czech Science Foundation and Ministry of Education, Youth and Sports of the Czech Republic, Hungarian National Research, Development and Innovation Office, New National Excellency Programme of the Hungarian Ministry of Human Capacities, Department of Atomic Energy and Department of Science and Technology of the Government of India, the National Science Centre and WUT ID-UB of Poland, the Ministry of Science, Education and Sports of the Republic of Croatia, German Bundesministerium f\"ur Bildung, Wissenschaft, Forschung and Technologie (BMBF), Helmholtz Association, Ministry of Education, Culture, Sports, Science, and Technology (MEXT) and Japan Society for the Promotion of Science (JSPS).
\end{acknowledgments}

\appendix
\section{}\label{app:a}
Here we provide a short summary of the use of generating functions in probability theory~\cite{probability_theory}. In the following we will assume that all random variables take only discrete non-negative values, i.e. $X \in \mathbb N$. 

Consider empirical probability distribution $p_{N}$ of the random variable $X$, say multiplicity distribution of protons, and construct power series in variable $w \in \langle0, 1\rangle$ with coefficients $p_{N}$.
\begin{equation}
Q(w) = \langle w^X \rangle = \sum_{N=0}p_{N}w^{N},
\end{equation}
where $\langle .\rangle$ indicates the average. Obviously knowing \textit{probability generating function} (pgf) $Q(w)$, we can retrieve $p_{N}$ from the derivatives $Q^{(N)}$ at $w = 0$
\begin{equation}
p_{N} = \left[ \frac{1}{N!}\frac{{\rm d}^NQ(w)}{{\rm d}w^N}\right]_{w=0} = \frac{Q^{(N)}(w=0)}{N!}.
\end{equation}

Similarly, the factorial moments are its derivatives at $w = 1$:
\begin{equation}\label{eq:a3}
\mu_{[r]} = Q^{(r)}(w=1)=\sum_{N\geq r}(N)_rp_{N}=\langle (X)_r\rangle,
\end{equation}

where $(N)_{r} = N(N-1) . . . (N+r-1)$ is falling factorial. Note that for multiplicity distributions $\mu_{[r]}$ represents the integral of the corresponding $r$-particle correlation functions.

Moreover, the pgf $Q(w)$ of the sum $X_1 + . . . + X_j$ of $j$ independent random variables equals the product of their pgfs $Q(w) = F_1(w) ... F_j (w)$. If the number of summands $j$ is a random variable with the pgf $G(w)$ and all $X_i$ are equally distributed with the pgf $F(w)$ then the pgf of their sum is compound function $Q(w) = G[F(w)]$.

Substitution $w = e^t$ into Eq. (A.1) leads to another set of generating functions.

\textit{Raw moment generating function} (rmgm) of random variable $X$ 
\begin{equation}\label{eq:a4}
M_X(t) = \langle e^{tX}\rangle = 1+\sum_{r\geq 1}\frac{\mu^{'}_rt^r}{r!}
\end{equation} is a power series in variable $t$ with the coefficients \mbox{$\mu_r^{'} = \langle X^r\rangle$}. Connection between the raw $\mu_r^{'}$ and factorial moments $\mu_{[r]}$ reads 
\begin{equation}
\begin{split}
\mu^{'}_r =& \sum_{j=0}^{j}S(r,j)\mu_{[j]}\\
=&\sum_{j=0}^{r}\frac{1}{j!}\sum^{j}_{i=0}(-1)^i\left( \begin{array}{c}j \\ i \end{array} \right) (j-i)^r\mu_{[j]},
\end{split}
\end{equation} where $S(r,j)$ are the Stirling numbers of the second kind~\cite{probability_theory}.

\textit{Central moments generating function} (cmgm)
\begin{equation}
\langle e^{(X-\mu)t}\rangle=e^{-\mu t}M_X=1+\sum_{r\geq 1}\frac{\mu_r t^r}{r!}
\end{equation} allows extraction of the moments $\mu_r = \langle (X-\mu)^r\rangle$ centered about the mean $\mu_1^{'}=\mu_{1}=\mu$.

\textit{Cumulant generating function} (cgf)
\begin{equation}\label{eq:a7}
K_X(t) = \ln M_X(t) = 1+ \sum_{r\geq 1}\frac{C_rt^r}{r!}
\end{equation} has coefficients $C_r$ called the cumulants at $t=0$.

The latter can be expressed via ordinary moments and vice versa~\cite{probability_theory,Smith1995ARF}
\begin{equation}\label{eq:a8}
\begin{split}
C_r &= \mu_r^{'} - \sum^{r-1}_{i=0}\left( \begin{array}{c}r-1 \\ i \end{array}\right) C_{r-i}\mu_{i}^{'}, \\ \mu_r^{'}&=\sum_{i=0}^{r-1}\left( \begin{array}{c}r-1 \\ i \end{array}\right) C_{r-i}\mu_i^{'}.
\end{split}
\end{equation}

With defining $s=e^{t}$, the \textit{factorial cumulant generating function} $K_f(s)$ is shown as
\begin{equation}
K_f(s)=\ln M_X(s).
\end{equation}
Then factorial cumulants $\kappa_r$ are derivatives of $K_f(s)$ at $s = 1$.
The derivatives of $s$ and $t$ are given by
\begin{equation} \label{eq:st}
\begin{split}
&\frac{\partial^r}{\partial t^r}s|_{s=1} = 1,\\
&\frac{\partial^r}{\partial s^r}t|_{t=0} = (-1)^{r-1}(r-1)!.
\end{split}
\end{equation} Using Eq.~\ref{eq:st}, cumulants and factorial cumulants can be expressed by each other.
A compact form is shown as 
\begin{equation}\label{eq:cfc}
\kappa^r = \langle N(N-1) ... (N-r+1)\rangle_c
\end{equation} where $\langle .\rangle_c \equiv C_r$~\cite{PhysRevC.95.064912}.

From the equality
\begin{equation}\label{eq:a9}
K_{X+a}(t) = at + K_X(t),\quad a=const.,
\end{equation} it follows that for $r\geq 2$ the coefficients of $t^r/r!$ in $K_{X+a}(t)$ and $K_{X}$(t) are the same. Moreover, Eq.~\ref{eq:a9} with $a=-\mu$ yields that first three cumulants and moments are equal $C_r=\mu_r$. Consequently, expressions for moments $\mu_r$ in terms of cumulants $C_r$ are obtained from Eq.~\ref{eq:a8} by dropping all terms with $C_1$.

From Eqs.~\ref{eq:a7} and \ref{eq:a4} it follows that cumulant of the sum $X_1+X_2$ of two independent random variables $X_1$ and $X_2$ is equal to the sum of the cumulant of $X_1$ and $X_2$.
\begin{equation}\label{eq:a10}
C_{X_1+X_2}(t) = \ln\langle e^{t(X_1+X_2)}\rangle = \ln \langle e^{tX_1}\rangle + \ln \langle e^{tX_2}\rangle.
\end{equation}

Consider superposition model of A+A collisions where the average proton multiplicity $\langle N_p\rangle$ as well as the cumulants $C_r$ are sums of contributions from elementary nucleon-nucleon scatterings and are therefore proportional to one another. Departure from linear behavior $C_r\sim\langle N_p\rangle$ in most central collisions occurs when $N_{\rm bin}$ comes to its limit. Its breakdown in non-central collisions revealed first in higher-order cumulants may signal transition to a regime dominated by strong multi-particle correlations.

\section{}\label{app:b}
Here we discuss some examples of \textit{Power series distributions} (PSDs) with pgf of the form $Q(w)=Z(\lambda w)/Z(\lambda)$. Their cumulants satisfy a simple recurrence relation~\cite{probability_theory}
\begin{equation}\label{eq:b1}
C_{r+1} = \lambda\frac{{\rm d}C_r}{{\rm d}\lambda}
\end{equation} making it possible to calculate higher-order cumulants starting from known $\mu_1=\mu_{[1]}=C_1$. It is also worth mentioning where $Z(\lambda, V, T)$ represents the grand canonical partition function at fixed fugacity $\lambda$, volume $V$, and temperature $T$.

\textit{Poisson distribution} (PD) 
\begin{equation}\label{eq:b2}
\begin{split}
p_N &= e^{-\lambda}\frac{\lambda^N}{N!}, \\Q(w)&=e^{(- \lambda)(w-1)},\\Z(\lambda)&=e^{-\lambda},\\\mu_{[r]} &= \lambda^r,\\C_r&= \lambda.
\end{split}
\end{equation}

\begin{equation}\label{eq:b3}
\begin{split}
p_N &= \left( \begin{array}{c} N+k-1 \\ k-1 \end{array}\right)(1-\lambda)^k\lambda^k,\\Q(w) &= \left( \begin{array}{c} 1-\lambda \\ 1-\lambda w \end{array}\right)^k,\\Z(\lambda)&=(1-\lambda)^{-k},\\\mu_{[r]}&=\frac{(k+r-1)!}{(k-1)!}\left( \begin{array}{c} \lambda \\ 1-\lambda \end{array}\right)^r.
\end{split}
\end{equation}

\textit{Negative binomial distribution} (NBD) with parameters $k > 0$ and $0 < \lambda < 1$ is an example of distribution which is over-dispersed to the Poisson distribution which is obtained as its limit $k \rightarrow \infty$, $\lambda \rightarrow 0$ with $k\lambda/(1-\lambda)$ fixed. 

Eq.~\ref{eq:b3} with $r=1$ yields $\mu_{[1]} = C_1=k\lambda/(1-\lambda)\equiv kx$. Pluggin $C_1$ into Eq.~\ref{eq:b1} yields
\begin{equation}
\begin{split}
C_2 &= C_1(1+x),\\
C_3 &= C_2(1+2x),\\
C_4 &= C_2(1+6x(1+x)),\\
C_5 &= C_2(1+2x)(1+12x(1+x)).
\end{split}
\end{equation}
Thus $C_r$ is a polynomial of order $r$ in $x$ with cumulant ratio $C_r/C_s$ independent of $k$.

\textit{Conway-Maxwell-Poisson distribution} (CMP)
\begin{equation}\label{eq:b5}
\begin{split}
p_N = \frac{\lambda^N}{(N!)^{\nu}}\frac{1}{Z(\lambda,\nu)},\\
Z(\lambda, \nu)=\sum_{N=0}\frac{\lambda^N}{(N!)^{\nu}}
\end{split}
\end{equation} with parameters $\lambda > 0, \nu > 0$ is used to model data which is either under-dispersed ($\nu>1$) or over-dispersed ($\nu < 1$) relative to the Poisson distribution ($\nu = 1$).

Expressions for factorial moments and cumulants of the CMP are rather cumbersome due to a complicated structure of the pgf $Z(\lambda w)/Z(\lambda)$, see Eq.~\ref{eq:b5}. Nevertheless, a simple formula generalizing the result for factorial moments of Poisson distribution (Eq.~\ref{eq:b2}) exists for its under-dispersed version with $r \in \mathbb N$~\cite{conway}:
\begin{equation}\label{eq:b6}
\langle ((N)_r)^{\nu}\rangle = \lambda^r.
\end{equation}

Our interest in CMP is motivated by the fact that for $\nu = 2$ it represents a stationary solution of the kinetic master equation describing the production of charged particles which are created or destroyed only in pairs due to the conservation of their charge~\cite{Ko:2000vp}. In this case Eq.~\ref{eq:b6} yields
\begin{equation}
\langle N^2\rangle = \mu_{{1}} + \mu_{[2]}=\lambda, \langle N^4\rangle=\lambda(1+\lambda),
\end{equation} which for $\lambda \gg 1$ leads to a factorization (de-correlation) of the raw moment $\langle N^4\rangle\approx\langle N^2\rangle\cdot\langle N^2\rangle$ connected to the underlying two-particle character of the charge correlations.

The same limit $\lambda \gg 1$ but for arbitrary $\nu$ was used in Ref.~\cite{conway} to obtain the asymptotic expression for the cumulants
\begin{equation}
C_r \approx \frac{\lambda^{1/\nu}}{\nu^{r-1}} + \mathcal{O}(1), \frac{C_r+1}{C_r}\approx\frac{1}{\nu}.
\end{equation}

Similarly to the case with $\nu=1$, $C_{r+1}/C_r$ becomes a $r$- independent constant which is bigger or smaller than one for $\nu<1$ or $\nu>1$, respectively.

\bibliography{main}% Produces the bibliography via BibTeX.

\end{document}